\documentclass[12pt,preprint]{aastex}

\begin{document}

\title{Triggering Collapse of the Presolar Dense Cloud Core 
and Injecting Short-Lived Radioisotopes with a Shock Wave. 
I. Varied Shock Speeds}

\author{Alan P.~Boss\altaffilmark{1}, Sandra A. Keiser\altaffilmark{1}, 
Sergei I. Ipatov\altaffilmark{1,2}, Elizabeth A. Myhill\altaffilmark{1,3}, 
and Harri A. T. Vanhala\altaffilmark{1,4}}
\altaffiltext{1}{Department of Terrestrial Magnetism, Carnegie Institution of
Washington, 5241 Broad Branch Road, NW, Washington, DC 20015-1305;
boss@dtm.ciw.edu, keiser@dtm.ciw.edu.}
\altaffiltext{2}{Catholic University, Washington, DC 20064; 
siipatov@hotmail.com.}
\altaffiltext{3}{Marymount University, Arlington, VA 22207;
elizabeth.myhill@marymount.edu.}
\altaffiltext{4}{NCESSE, Washington, DC 20791; 
HarriVanhala@ncesse.org.}

\begin{abstract}

 The discovery of decay products of a short-lived radioisotope (SLRI) in the 
Allende meteorite led to the hypothesis that a supernova shock wave 
transported freshly synthesized SLRI to the presolar dense cloud core, 
triggered its self-gravitational collapse, and injected the SLRI 
into the core. Previous multidimensional numerical calculations of the 
shock-cloud collision process showed that this hypothesis is plausible 
when the shock wave and dense cloud core are assumed to remain isothermal at
$\sim 10$ K, but not when compressional heating to $\sim 1000$ K is assumed. 
Our two-dimensional models (Boss et al. 2008) with the FLASH2.5 adaptive 
mesh refinement (AMR) hydrodynamics code have shown that a 20 km/sec
shock front can simultaneously trigger collapse of a 1 $M_\odot$ core
and inject shock wave material, provided that cooling by molecular species 
such as H$_2$O, CO, and H$_2$ is included. Here we present the results 
for similar calculations with shock speeds ranging from 1 km/sec to
100 km/sec. We find that shock speeds in the range from 5 km/sec to 
70 km/sec are able to trigger the collapse of a 2.2 $M_\odot$ cloud
while simultaneously injecting shock wave material: lower speed shocks 
do not achieve injection, while higher speed shocks do not trigger 
sustained collapse. The calculations continue to support the shock-wave
trigger hypothesis for the formation of the solar system, though the
injection efficiencies in the present models are lower than desired.

\end{abstract}


\section{Introduction}

 Triggering the collapse of the presolar cloud with an interstellar shock 
wave propagating away from a site of stellar nucleosynthesis is a favored
explanation for the widespread evidence of short-lived radioisotopes (SLRI) 
in chondritic refractory inclusions (Lee et al. 1976; Cameron and Truran 
1977; MacPherson, Davis and Zinner 1995) and, much more rarely, in chondrules 
(Russell et al. 1996) found in primitive meteorites. The goal of this
paper is to continue the theoretical exploration of the trigggering and
injection scenario for SLRIs, in the context of shock waves striking a
dense molecular cloud core that could have collapsed to form the solar
system.

\subsection{Short-Lived Radioisotopes}

 The dozen or so confirmed ($^{10}$Be, $^{41}$Ca, $^{26}$Al, $^{60}$Fe, 
$^{53}$Mn, $^{107}$Pd, $^{182}$Hf, $^{129}$I, and $^{244}$Pu)
or suspected ($^{99}$Tc, $^{36}$Cl, $^{205}$Pb, and $^{92}$Nb) 
short-lived radioisotopes may require a fairly involved history 
for their complete explanation (Goswami and Vanhala 2000;
Meyer and Clayton 2000; McKeegan \& Davis 2003; Wadhwa et al. 2007). 
A stellar nucleosynthetic source (a supernova or 
an AGB star) has been the leading explanation for most of these nuclei
(Cameron 1993, 2001; Wasserburg et al. 1994, 1995, 1998;
Trigo-Rodr\'iguez et al. 2009; Huss et al. 2009), though
there are other possibilities as well, in particular local production 
(i.e., in the solar nebula) of short-lived radioisotopes 
produced during spallation reactions involving energetic 
particles emanating from protosolar flares (Shu et al. 1997).
Such local irradiation models appear to have a problem with being 
able to match the observed abundance ratio of $^{26}$Al to $^{41}$Ca 
(Srinivasan et al. 1996; Sahijpal et al. 1998; 
Lee et al. 1998). The observed abundance of $^{26}$Al thus seems
to require its production by stellar nucleosynthesis (McKeegan 
et al. 2000). However, this problem can be avoided by assuming 
that the refractory inclusions are shielded by a less refractory mantle
during the irradiation, with the mantle being lost later on during heating 
of the inclusions, thereby yielding the approximate abundance ratio of
$^{26}$Al to $^{41}$Ca observed in certain meteorites 
(Gounelle et al. 2001). On the other hand, producing $^{26}$Al 
by multiple episodes of local irradiation would negate its use as a 
precise chronometer for the early solar system (Bizzarro et al. 2004; 
Halliday 2004; Krot et al. 2005; Thrane et al. 2006), which 
seems to be ruled out by the agreement of $^{26}$Al ages with those 
derived from the Pb-Pb dating system (Connelly et al. 2008).

 Evidence has also appeared for the 
presence of the short-lived isotope $^{10}$Be in an Allende inclusion 
(McKeegan et al. 2000). Because $^{10}$Be is thought to be 
produced only by nuclear spallation reactions, its existence 
has been used to argue strongly in favour of local irradiation
(McKeegan et al. 2000; Gounelle et al. 2001).
Sahijpal \& Gupta (2009) have calculated that even if all of the
$^{10}$Be was produced by local irradiation, then the amount of $^{26}$Al
also produced by local irradiation was about 10\% of the total amount
of $^{26}$Al, so that the bulk of the $^{26}$Al was probably
synthesized by a massive star. 
However, if irradiation is responsible for the $^{10}$Be,
it is unclear if the irradiation occurred in the solar nebula,
or in an earlier phase of evolution. Arnould et al. (2000)
point out that spallation can occur in the winds ejected from
H-depleted Wolf-Rayet (WR) stars. Desch et al. (2004) showed
that the $^{10}$Be might well have originated from $^{10}$Be galactic
cosmic rays that were stopped in the presolar cloud. Evidence
has also been advanced for the presence of live $^7$Be in
Ca,Al-rich, refractory inclusions (CAIs), which are believed to represent
the earliest solids formed in the solar nebula that have survived
relatively unaltered (Chaussidon et al. 2006). Because of the
extremely short half-life of $^7$Be of 53 days, this evidence, if correct,
would require formation of $^7$Be in the solar nebula. Desch \& 
Ouellette (2006) have disputed the $^7$Be claim, which has not
been confirmed by other groups to date.

 The short-lived isotope $^{60}$Fe (Goswami and Vanhala 2000) cannot be 
produced in the appropriate amounts by spallation, and requires a 
stellar nucleosynthetic source (Tachibana \& Huss 2003), as does the bulk
of the $^{26}$Al observed to be polluting the interstellar medium.
The half-life of $^{60}$Fe is 2.6 million years (Rugel et al. 2009), 
roughly four times 
that of $^{26}$Al, so in any case, the evidence for live short-lived
isotopes in refractory inclusions seems to require that no more 
than about 1 million years elapsed between the nucleosynthesis of some of
the short-lived isotopes in a star and the formation of refractory
inclusions in the solar nebula. 

 Solar-type stars are believed to form from the collapse of dense 
molecular cloud cores, which are supported against collapse primarily by 
magnetic fields (e.g., Mouschovias, Tassis, \& Kunz 2006),
though turbulence also plays a role (e.g., Kudoh \& Basu 2008).
Collapse of a quiescent cloud core begins once the magnetic field support
decreases sufficiently through the process of ambipolar diffusion.
Recently Kunz \& Mouschovias (2009) have shown that ambipolar diffusion
leading to collapse and fragmentation is able to reproduce the
observed distribution of molecular cloud core masses, i.e., the initial
core mass function, suggesting the importance of magnetic fields
for star formation in general. Ambipolar diffusion is estimated to 
require of order 10 Myr to lead to collapse (Mouschovias, Tassis, 
\& Kunz 2006), a period considerably longer than the half-life 
of $^{26}$Al of 0.73 Myr. If the Solar System's $^{26}$Al was produced 
in a massive star, it may have been injected promptly into the protosolar
cloud, which must have then collapsed and formed cm-sized solids, all
within $\sim 1$ Myr. This constraint assumes that the same stellar
outflow that carried the $^{26}$Al may have triggered the collapse of 
the protosolar cloud and injected other newly-synthesized elements, 
including other short-lived isotopes (Cameron \& Truran 1977;
Boss 1995). Abundant observational support exists 
for the triggering of star formation by expanding supernova shells 
in Upper Scorpius (Preibisch \& Zinnecker 1999; Preibisch et al. 2002) 
and the Cygnus Loop (Patnaude et al. 2002), by superbubbles in OB 
associations (Oey et al. 2005; Lee \& Chen 2009), by ionization fronts 
associated with HII regions (Leppanen, Liljestrom, \& Diamond 1998;
Healey, Hester, \& Claussen 2004; Hester \& Desch 2005; Snider et al. 
2009), by generic external shocks (Tachihara et al. 2002), and by 
protostellar outflows (Barsony et al. 1998; Sandell \& Knee 2001; 
Yokogawa et al. 2003). Here we consider generic shocks, with a special
emphasis on supernovae and AGB stars as shock sources.

\subsection{Injection Scenarios}

 Boss (1995) showed that shock fronts from a nearby AGB star or a 
relatively distant supernova could trigger the collapse of a 3D dense 
cloud core and inject shock front material into the collapsing cloud. 
Foster \& Boss (1996, 1997) studied this process in greater
detail for axisymmetric, 2D clouds, and pointed out the crucial
role of the assumed isothermal shock front for achieving both
goals of triggered collapse and injection. 
A supernova shock passes through three phases: ejecta-dominated,
Sedov blast wave, and radiative (Chevalier 1974). The latter phase 
occurs at distances of about 10 pc, after which the shock front 
sweeps up a cool shell of gas and dust as it propagates.
Several recent AMR studies (Nakamura et al. 2006; 
Melioli et al. 2006) have confirmed the results of
Boss (1995) and Foster \& Boss (1996, 1997) that shock-triggered
star formation is likely to occur when the supernova
shock front has evolved into a radiative shock, i.e., the
shock front is able to cool so rapidly by radiation that the thin shock front
gas is essentially at the same temperature as the ambient gas, which is the
same situation as in the isothermal shocks considered by
Boss (1995) and Foster \& Boss (1996, 1997).

 Rayleigh-Taylor (R-T) fingers were identified as the 
physical mechanism for achieving injection of dust grains and 
gas into the collapsing presolar cloud (Foster \& Boss 1997; 
Vanhala \& Boss 2000, 2002). Because the R-T fingers strike the
outermost layers of the presolar cloud, inducing collapse, the
R-T fingers do not reach the central regions until shortly after
the central protosun and the early solar nebula have formed,
possibly explaining the lack of $^{26}$Al in certain (Fraction
Unknown Nuclear = FUN) refractory
inclusions  (Sahijpal \& Goswami 1998) that may have formed before 
the R-T fingers arrived. Boss (2007) modeled the R-T finger 
injection process in the context of this scenario simply by imagining 
spraying the $^{26}$Al onto the surface of an existing solar nebula.

 A related but alternative scenario involves having a 
nearby ($\sim$ 0.1 pc) supernova inject $^{26}$Al directly 
into the solar nebula (rather than into the presolar
cloud), as studied by Ouellette, Desch \& Hester (2007). They 
found that the gas from a shock front could not be injected efficiently
into a protoplanetary disk because of the disk's much higher density
compared to the presolar cloud. Chevalier (2000) had found the same 
result and attributed it to the hot shock gas not having enough time 
to cool down. As a result, Ouellette et al. (2007) suggested
that the $^{26}$Al resided primarily in dust grains that shot
through the stalled shock-front gas and thereby penetrated into the 
disk. Dust grains smaller than 0.1 micron would be deflected, but
micron-sized and larger dust grains would be injected with a high 
efficiency (Ouellette et al. 2009). 

 Supernovae are known to produce
large amounts of dust grains, but theoretical models suggest that
the newly condensed grains are essentially all smaller than 0.1 micron, 
and are sputtered to even smaller sizes in the reverse shock front 
driven into the expanding supernova remnant (SNR) by the interstellar
medium that the SNR encounters (Bianchi \& Schneider 2007). The models
are in accord with dust extinction estimates for an observed reddened QSO.
An insignificant fraction of the total dust grain mass is contained 
in grains larger than 0.1 micron, presenting a problem for scenarios
that rely on large dust grains for injection. However, Nittler (2007) 
has argued that a sub-class of presolar grains appears to have been 
formed in a single supernova, conceivably the same supernova that produced
many SLRIs. These presolar grains have sizes of 0.1 to 10 micron,
large enough to be injected into the disk, or the presolar cloud core.
What fraction of the mass of the initial dust grain population these
relatively large grains represent is difficult to determine, 
given the processing associated with detecting presolar grains in 
meterorites and the typical limitation to the study of grains larger
than about 0.1 micron (e.g., Amari, Lewis, \& Anders 1994).

 Recently it has been suggested that $^{60}$Fe was injected by
a supernova directly into the solar nebula roughly 1 Myr after 
Solar System formation (Bizzarro et al. 2007), as in the 
Ouellette et al. (2007) models. In this scenario, the $^{26}$Al 
was derived from the wind from a Wolf-Rayet (WR) star, a massive star
that would later become a supernova and inject the $^{60}$Fe.
WR winds do indeed carry large amounts of $^{26}$Al (Arnould,
Goriely, \& Meynet 2006), with winds that are comparable in
speed ($\sim$ 1500 km/sec) to supernova shocks (Marchenko et al.
2006), meaning that such winds would need to be slowed down
by sweeping up interstellar gas and dust to speeds less than
$\sim 40$ km/sec if they are to trigger cloud collapse rather
than simply shred clouds to pieces (Foster \& Boss 1996, 1997).
However, other groups have not been able to replicate the
Ni isotope data that forms the basis for the $^{60}$Fe scenario
(Dauphas et al. 2008; Regelous et al. 2008). Nevertheless, WR stars
should be considered as a possible source of $^{26}$Al in
addition to that obtained from a supernova (e.g., 
Gaidos et al. 2009).

 Williams \& Gaidos (2007) estimated that the likelihood of a 
protoplanetary disk being struck by a supernova shock was less 
than 1\%, but considering that we do not know if the Solar System's
inventory of short-lived isotopes is rare or not, such an argument 
cannot be considered decisive. In fact, it has been argued
recently that significant $^{26}$Al is necessary for the development
of technological civilizations (Gilmour \& Middleton 2009).
However, Gounelle \& Meibom (2007) and 
Krot et al. (2008) argued that injection could not have occurred 
directly onto a relatively late-phase, low mass solar nebula,
as $^{26}$Al from a massive star supernova would have been accompanied 
by sufficient oxygen to lead to an oxygen isotope distribution
in the solar nebula that would be distinct from that inferred for 
the sun based on mass-independent fractionation of carbonaceous chondrites 
and from that recently measured by the Genesis Mission (McKeegan
et al. 2008). Krot et al. (2008) therefore argued that injection must 
have instead occurred into the presolar cloud, so that the sun and 
the solar nebula shared a common reservoir of oxygen isotopes that
could then undergo fractionation. Ellinger, Young, \& Desch (2009),
however, pointed out that supernova explosions are not spherically
symmetric, and so some degree of anisotropy in the ejecta is to be
expected, perhaps enough to permit $^{26}$Al injection into the
presolar nebula without disturbing the oxygen isotope ratios.

\subsection{Shock Thermodynamics}

 While isothermal shock fronts are capable of simultaneous triggering 
and injection (Boss 1995; Foster \& Boss 1996, 1997; Vanhala \& Boss
2000, 2002), it has been less clear what happens when detailed 
heating and cooling processes in the shock
front are considered. Vanhala and Cameron (1998) found that when 
they allowed nonisothermal shocks in their models, they could
not find a combination of target cloud and shock wave parameters
that permitted both triggered collapse and injection to occur:
they could trigger cloud collapse, or they could inject particles,
but not both in the same simulation. Such an outcome would be fatal 
to the triggering and injection hypothesis if definitive. However, 
Vanhala \& Cameron's (1998) models employed a smoothed-particle 
hydrodynamics (SPH) code, which has since been shown to be poor at 
resolving dynamical instabilities such as the Rayleigh-Taylor or 
Kelvin-Helmholtz instabilities (Agertz et al. 2007). Furthermore,
Vanhala and Cameron's (1998) thermodynamical routines led to post-shock 
thermal profiles that were quite different from those of Kaufman \& 
Neufeld (1996), who found that for a 40 km/sec shock, the post-shock gas 
cooled down from a maximum of over 3000 K to less than 100 K within a 
distance of 0.001 pc. With a 25 km/sec shock in the Vanhala \& Cameron
(1996) SPH code, however, the post-shock temperature rose to 3000 K 
and showed no signs of decreasing over a distance on the order of 0.5 pc. 

 Kaufmann \& Neufeld (1996) studied C-type shock fronts, which 
result in the most successful models of the shock emission from the 
Kleinmann-Low nebula in the Orion molecular cloud. 
Kaufman \& Neufeld (1996) assumed preshock magnetic field strengths
in their C-shock models that are consistent with magnetic field strengths 
measured by Zeeman splitting in molecular clouds (Crutcher 1999). Such
nondissociative, magnetohydrodynamic C-type shocks appear 
to be the correct analogue for the shock speeds 
($\sim$ 5 to 40 km s$^{-1}$) that we expect will be necessary to
simultaneously achieve triggered collapse and injection. The
relatively low postshock temperatures in C-type shocks are crucial
for this scenario. Postshock cooling depends sensitively on 
the detailed microphysics, e.g., on the emission from rotational
states of H$_2$O, H$_2$, CO, and OH, and hence on 
quantities such as the ratio of ortho- to para-hydrogen molecules
(Neufeld \& Kaufmann 1993; Kaufmann \& Neufeld 1996). Atomic species 
are also important coolants, as are dust grains. More recently
Morris, Desch, \& Ciesla (2009) have reexamined the question of
cooling shocked gas by H$_2$O line emission, finding that the
cooling rates in the optically thin limit are at least as
high as those calculated by Neufeld \& Kaufmann (1993).

 Boss et al. (2008) found that simultaneous triggering and
injection was possible for a 20 km/sec shock striking a 1 $M_\odot$
cloud, provided that Neufeld \& Kaufmann's (1993) molecular cooling 
functions were employed. Gounelle et al. (2009) interpreted this result 
as meaning that only a very narrow range of shock speeds was consistent 
with the triggering and injection hypothesis. In this paper
we examine what happens
for a wide range of shock speeds (and different mass clouds) compared
to the single shock speed considered by Boss et al. (2008), in order 
to determine the robustness of the shock wave trigger hypothesis.

\section{Numerical Methods}

 Achieving adequate spatial resolution of shock-compressed
regions of presolar clouds hit with supernova shock fronts, while
minimizing the overall computational burden, is an insurmountable
problem for a fixed grid hydro code of the type previously
used on these problems (Boss 1995; Foster \& Boss 1996, 1997; 
Vanhala \& Boss 2000, 2002). Clearly the demands of the shock-triggered 
collapse and injection problem requires adoption of the Adaptive Mesh 
Refinement (AMR; e.g., Truelove et al. 1997; Poludnenko 
et al. 2002) technique, which was designed to handle just this 
type of situation. AMR techniques automatically
insert new grid points in regions of strong physical gradients, and
remove them in regions without strong gradients, in order to
maximize the spatial resolution in the crucial regions while
minimizing the computational burden.

 FLASH employs a block-structured adaptive grid approach using the 
PARAMESH package. Advection is handled by the piecewise parabolic method 
(PPM), which features a Riemann solver at cell boundaries that handles 
shock fronts exceptionally well. In FLASH, PPM is incorporated in a 
form that is second-order accurate in space and time. 
We have tested the FLASH2.5 code's ability to reproduce
the results of several different test cases that are relevant to
the problem of triggering cloud collapse, namely the Sod shock
tube problem and the collapse of a pressureless sphere. While
the performance of FLASH on the pressureless sphere collapse is
not as accurate as with codes designed to study collapse problems
(e.g., Boss \& Myhill 1992), FLASH does a superb job of handling the 
Sod shock tube problem on a Cartesian grid, when the shock flows 
parallel to one axis or at a 45 degree angle.
While the standard FLASH test cases were run on Cartesian grids,
we have also reproduced the correct results for the Sod shock
tube and for pressureless cloud collapse on the cylindrical coordinate
($R,Z$) grid that is used in the present calculations.
  
 In the absence of cooling or an isothermal constraint, FLASH 
produces an adiabatic evolution with an effective $\gamma = 5/3$. 
The FLASH equation of state routines were taken to be those for a 
simple perfect gas with a mean molecular weight of $\mu = 2.3$.
We have adapted these FLASH routines to simulate isothermal
shock-cloud interactions, where the entire computational grid
is forced to remain isothermal, typically at 10 K, in order to
compare with the results of Foster \& Boss (1996). In these
models, $\gamma$ is set equal to 1.01, as a value of $\gamma = 1.0$
is prohibited by the Riemann solver.  

 In addition to isothermal models, we also present models 
that employed the same compressional heating and 
radiative cooling that was studied by Boss et al. (2008). Our model 
for radiative cooling is based on the results of Neufeld \& Kaufman (1993),
who calculated the radiative cooling caused by rotational
and vibrational transitions of optically thin, warm molecular gas
composed of H$_2$O, CO, and H$_2$ and found H$_2$O to be the
dominant cooling agent. Neufeld \& Kaufman's (1993) Figure 3 shows 
that over the range of temperatures from 100 K to 4000 K, the total
cooling rate coefficient $L$ can be approximated as
$L \approx L_0 \approx 10^{-24} (T/100)$ erg cm$^3$ s$^{-1}$.
The cooling rate $\Lambda = L \ n(H_2) \ n(m)$, where $n(H_2)$
is the number density of hydrogen molecules and $n(m)$ is
the number density of the molecular species under consideration.
Assuming that $n(H_2O)/n(H_2) \approx 8.8 \times 10^{-4}$
(Neufeld \& Kaufman 1993), we take $n(m)/n(H_2) \approx 10^{-3}$, 
leading to a radiative cooling rate of
$\Lambda \approx 9 \times 10^{19} (T/100) \rho^2$ erg cm$^{-3}$ s$^{-1}$,
where $\rho$ is the gas density in g cm$^{-3}$. Boss et al. (2008)
found that $\Lambda$ could be increased or decreased by factors
of two without having a major effect on the outcome of shock
triggering and injection, so the precise value of $\Lambda$ does
not appear to be critical to the results. 

 Kaufman \& Neufeld (1996) found that the peak temperatures in their
MHD shocks were typically of order 1000 K for shock speeds in the
range from 5 km/sec to 45 km/sec. As a result, the
temperatures in the present models as well as those of Boss et al. 
(2008) have been restricted to values between 10 K and 1000 K.
Kaufman \& Neufeld (1996) also found a typical shock thickness of order
0.001 pc = $3 \times 10^{15}$ cm for a 40 km/sec shock propagating
in a magnetized gas with a preshock density similar to that of 
dense cloud cores. A similar shock thickness occurs in the models
of Boss et al. (2008), using the Neufeld \& Kaufman (1993) radiative
cooling rate $\Lambda$, though with a 20 km/sec shock speed and
an unmagnetized cloud: evidently the higher shock speed in
Kaufman \& Neufeld (1996) is roughly compensated for by the
presence of magnetic fields, compared to the Boss et al. (2008)
results.

 As in Boss et al. (2008), in the present models we used the two 
dimensional, cylindrical coordinate ($R, Z$) version of FLASH2.5, 
with axisymmetry about the rotational axis ($\hat z$). Multipole 
gravity was used, including Legendre polynomials up to $l = 10$. 
The cylindrical grid is typically 0.197 pc long in $Z$ and 0.063 pc
wide in $R$, though in the higher shock speed models, the grid was 
extended to be 0.320 pc long in order to better follow the interaction. 
We set the number of blocks in $R$ ($N_{BR}$) to be 5 in all cases,
while the number of blocks in $Z$ ($N_{BZ}$) has been varied from 
5 to 20. Most models have had 15 blocks in $Z$, leading to an 
approximately uniform grid spacing in $R$ and $Z$. With each 
block consisting of $8 \times 8$ grid points, this is equivalent 
to an initial grid of $40 \times 120$ for most models. The number
of levels of refinement ($N_L$) has been varied from five to six.
With five levels of refinement employed, FLASH is
able to follow small-scale structures with the
effective resolution of a grid 16 times finer in scale, or
effectively $640 \times 1920$, somewhat better than the highest 
resolution of $480 \times 1440$ used by Vanhala \& Boss (2000).
With six levels, the resolution is increased by another factor
of two in each direction.

\section{Initial Conditions}

 Our target dense cloud cores consist of Bonner-Ebert (BE) spheres
(Bonnor 1956), which are the equilibrium structures for self-gravitating,
isothermal spheres of gas. BE spheres are excellent models for the 
structure of pre-collapse dense molecular cloud cores seen in star-forming 
regions (e.g., Shirley et al. 2005). For the models that compare 
results to the isothermal models of Foster \& Boss (1996, 1997), the 
target dense cloud core is a BE sphere with a mass of 1.1 $M_\odot$, a 
radius of 0.058 pc, a temperature of $T = 10$ K, and a maximum density of 
$6.2 \times 10^{-19}$ g cm$^{-3}$, located at rest near the top of
the cylindrical grid. The BE sphere is embedded in an intercloud
medium with a density of $3.6 \times 10^{-22}$ g cm$^{-3}$ 
and a temperature of 10 K. The shock wave begins at the top of
the grid and propagates downward at a specified speed toward the
BE sphere. The shock wave has a thickness of 0.003 pc with a 
uniform density of $3.6 \times 10^{-20}$ g cm$^{-3}$, a mass of
0.016 $M_\odot$, and a temperature of 10 K. 
For the models where cooling is included, the shock
wave begins with a temperature of 1000 K and is followed by a
post-shock wind with a density of $3.6 \times 10^{-22}$ g cm$^{-3}$ 
and temperature of 1000 K, also moving downward at the same speed
as the shock wave.

 The assumed shock structure is the same 
as that used in the standard case of Foster \& Boss (1996, 1997)
and investigated by Boss et al. (2008). The shock structure was chosen
to resemble the expected conditions in a planetary nebula wind (e.g.,
Plait \& Soker 1990; see discussion in Foster \& Boss 1996).
However, it is also consistent with a supernova shock that has
swept-up considerable material and slowed down as a result.
Chevalier (1974) considered a supernova shock propagating
into a cold (T = 10 K) medium with a number density of 1 cm$^{-3}$.
By 0.25 Myr, the shock has slowed to a top speed of $\sim$ 60 km/sec
and has travelled a distance of $\sim$ 2.5 pc. The amount of
swept-up mass contained in the shock front that is incident on
the target clouds used in the present calculations is 0.015 $M_\odot$,
quite close to the value of 0.016 $M_\odot$ in the standard case.

 The shock wave material is represented by a color field, initially 
defined to be equal to 1 inside the shock wave and 0 elsewhere,
which allows the shock wave material to be tracked in time 
(Foster \& Boss 1997). The SLRI are assumed to be contained primarily
in dust grains of sub-micron size (e.g., Bianchi \& Schneider 2007), 
small enough for the grains to remain coupled to the gas. Grains
larger than this size could shoot through the shock front as it
strikes the target cloud and increase the injection efficiency
(e.g., Ouellette et al. 2007, 2009), so injection efficiencies
derived solely from the color field approach should be considered as
lower bounds on the true injection efficiencies.

\section{Results}

 We present results for several related studies with the FLASH code,
namely a set of comparisons with the results on the standard case of
Foster \& Boss (1996, 1997), the stability of BE-like spheres 
with higher central densities (and higher masses) in the absence 
of shock waves, and, finally, the effects of varying the shock
wave speed across a wide range of values for a BE-like target 
sphere with a mass of $2.2 M_\odot$.

\subsection{Standard Case Comparisons}

 We first used FLASH2.5 to reproduce the standard case of triggered 
isothermal collapse of Foster \& Boss (1996), and verified that 
FLASH2.5 was able to produce simultaneous triggered collapse and 
injection of shock wave material in the case of a 20 km/sec shock.

 Table 1 lists four models that duplicate the standard case,
with varied spatial resolution in FLASH2.5. Table 1 also gives
the results of the models: the maximum density obtained ($\rho_{max}$) 
in g cm$^{-3}$, the fraction of the incident color field that is 
injected ($f_i$), and the final time of the model ($t_f$) in seconds.
Once maximum densities of $\sim 10^{-12}$ g cm$^{-3}$ are reached,
even with five or six levels of refinement, FLASH is unable to
follow the collapse to even higher densities, either because it
does not have sufficient spatial resolution on the scale of
the density maximum, which typically occupies only a few grid
cells, or else because the Poisson solver is unable to properly
represent the gravitational potential of what has become in
essence a point mass. As a result, the final times listed in Table 1
do not correspond to the time when the density maximum is reached, 
as FLASH simply continues to do its best to evolve the entire
cloud-shock system past that instant of time. Once maximum
densities of $\sim 10^{-12}$ g cm$^{-3}$ are reached, it is
appropriate to terminate these calculations anyway, because by
such densities the collapsing cores have become optically thick
and can no longer cool by our assumed molecular cooling law,
which assumes optically thin clouds. A full radiative
transfer treatment (e.g., Boss \& Myhill 1992) is required
to treat optically thick regions, a capability that does not
exist in FLASH2.5. Finally,
as in Boss et al. (2008), $f_i$ is defined to be that fraction
of the initial color field that is incident on the initial target
cloud and that is injected into regions of the collapsing cloud 
core with density greater than $10^{-18}$ g cm$^{-3}$. That is
to say, if the total amount of color that is incident on
the target cloud is 1 in dimensionless units, $f_i = 0.001$
means that 0.001 is the amount of color that was injected.

 It is clear from Table 1 that in all four models, the BE sphere
was triggered into self-gravitational collapse, given that 
maximum densities of $\sim 10^{-12}$ g cm$^{-3}$ or higher were
achieved in each case. This maximum density is well over a factor
of $10^6$ times higher than the initial maximum density in the
target BE sphere of $6.2 \times 10^{-19}$ g cm$^{-3}$, indicating
that dynamical collapse has been induced in these cloud cores.
It is also clear that as the spatial resolution in FLASH is
increased, the collapsing cloud is able to reach slightly
higher densities, implying that the highest resolution calculations
are approaching the continuum limit. Note also that the injection 
efficiency obtained for model FBD, 0.002, is only slightly lower than
the efficiency of 0.003 obtained for the 20 km/sec shock model with 
heating and cooling presented by Boss et al. (2008), which had 
the same spatial resolution as model FBD, showing that when
nonisothermal processes are considered, the injection process becomes
only somewhat less efficient than when isothermality is assumed.

 The evolution of these four isothermal models is very similar to that 
of the models to be presented below with compressional heating and 
radiative cooling, and so separate figures for the isothermal
models are not displayed.

\subsection{Stability of Varied Mass Spheres}

 We have also used FLASH to verify the long-term stability of the
target clouds in the absence of a triggering shock front, as clouds
that collapse on their own on time scales similar to the shock wave
passage cannot be considered to have been triggered into collapse.
Shocked-triggered collapse for the standard case occurs within a time 
span of about $10^5$ yrs (Foster \& Boss 1996, 1997; Boss et al. 2008).
When the target BE sphere used in the models in Table 1 
is evolved isothermally in FLASH2.5 without being struck by a shock wave, 
it does not collapse, but instead oscillates around its initial
equilibrium structure, over a time period of at least $10^6$ yrs,
sufficiently long to validate the claim that triggering has occurred
in the Foster \& Boss (1996) comparison models.

 In addition, we have tested the stability of higher mass cloud
cores, obtained simply by multipying the initial densities
of the standard case BE sphere by factors of 2, 3, or 4. 
Formally speaking, these clouds are not true BE spheres in
equilibrium, but any deleterious effects 
of this simple computational convenience
can be discerned by seeing if the clouds do indeed collapse on
their own without being subjected to shock triggering.
In the case of the two clouds where the densities had been
increased by factors of 3 or 4, the clouds did indeed begin
to undergo dynamic collapse to densities of
$\sim 10^{-12}$ g cm$^{-3}$ or higher within times of
$\sim 10^5$ yrs, showing their gravitational instability
and unsuitability for the shock-triggering models.
However, the cloud with twice the BE sphere density, i.e.,
an initial central density of $1.24 \times 10^{-18}$ g cm$^{-3}$
and a mass of 2.2 $M_\odot$, remained stable in FLASH for at 
least $\sim 10^6$ yrs, proving its suitability for the 
present studies. The next section then presents the results
for when a 2.2 $M_\odot$ cloud core is struck by shocks
with varying speeds. 

\subsection{Varied Shock Wave Speeds}

 We now turn to the main focus of this paper, a consideration of
what happens when shock waves with a wide variety of speeds are
allowed to strike a dense cloud core, while including a fully
nonisothermal treatment with compressional heating and radiative
cooling.

 Table 2 lists the shock speeds employed for these models of a
2.2 $M_\odot$ target cloud core, as well as the results of
the calculations, as in Table 1. Several variations on the spatial
resolution were also calculated: models v4-4 and v5-4 had four levels
of adaptive mesh refinement, rather than five, while model v20-6
had six. The higher speed models (v75L, v80L, v90L, and v100L) had
longer grids in the $Z$ (vertical) direction, in order to better
follow the evolution of the cloud as it is swept downward by the
shock front. Figure 1 depicts the initial density distribution for
all of the nonisothermal models with varied shock speeds, showing
the target cloud and the incoming shock front, where the color
field is taken to be initially uniform with a value of unity.

 Figures 1 through 3 show the evolution of model v20, which 
typifies the results for models resulting in simultaneous collapse
and injection. Model v20 is also identical to model C of Boss et al. 
(2008), except for having a cloud mass of 2.2 $M_\odot$ instead of 
1 $M_\odot$, and behaves in a very similar manner, implying that
the process works equally well for target clouds in this mass range.

 Figure 2 shows that the model v20 shock front is able to compress 
the top edge of the target cloud, while Rayleigh-Taylor (R-T)
fingers and Kelvin-Helmholtz (K-H) vortices form around the
shock-cloud interface. The R-T fingers drive into the
target cloud, while the K-H vortices tend to ablate material
off the cloud edge and force it to join the downstream flow.
The contours for the color field in Figure 2 show that most of
the color field impinging on the target cloud is diverted
from the cloud by these vortices and disappears downstream.
Figure 4 shows a close-up of the shock-cloud interface of
model v20 at the same time as Figure 2, providing a better
look at the structure of the R-T and K-H features, as well
as of the color field. 

 Figure 3 shows that model v20 is able to form a dense clump
along the symmetry axis less than 0.1 Myr after the evolution
began. Figure 5 shows a close-up of the dense clump at the
same time as Figure 3, this time with the temperature field
contoured in black, instead of the color field. Temperatures
rise above 100 K only in regions immediately adjacent to the
shock-cloud interface, i.e., in the region with the largest
density gradients and hence the strongest compressional
heating. Except for this highly shocked region, then, the
bulk of the cloud is able to remain nearly isothermal at
the assumed background temperature of 10 K as a result of the
cooling by molecular species such as H$_2$O, CO, and H$_2$.
The strength of this molecular cooling explains why the
results of the nonisothermal model presented by Boss et al. (2008)
are so similar to those of the isothermal standard test case of
Foster \& Boss (1996, 1997).

 Figure 6 shows an even closer-in view of the density maximum
of model v20 after a time of 0.1 Myr. The velocity vectors
indicate that the region in the vicinity of the density
maximum is trying to collapse onto the clump with speeds
as high as several km/sec, which is highly supersonic,
considering that the sound speed in 10 K gas is 0.2 km/sec.
Evidently, a protostar with a maximum density of $\sim 10^{-13}$
g cm$^{-3}$ has formed and is growing by the accretion of
gas from the target cloud's envelope. Meanwhile, the protostar
has been accelerated by the shock front to a speed of order
1 km/sec, and is moving downward as a result. Figure 7 plots
the color field over the same region as Figure 6, showing that
significant shock front material has been injected into the 
collapsing protostar and the infalling cloud envelope. 
Multiple waves of shock front material should
be accreted by the protostar, considering the velocity field
evident in Figure 7, including the regions with even higher color
densities than that within the density maximum, which appear likely
to collapse onto the protostar within a few thousand more years.

 Figure 8, 9, and 10 show the evolution of model v4, where dynamic collapse
was triggered, but no significant injection occurred. Comparing
Figure 8 for model v4 with Figure 2 for model v20, at comparable
phases of shock-cloud interaction, it is apparent that the
much slower speed shock front in model v4 is unable to compress
the target cloud's edge to the extent achieved by the shock
in model v20. In fact, by a time of 0.105 Myr (Figure 9), the
cloud in model v4 has been clearly triggered into collapse by
the shock front, yet the color field lags behind and seems
unlikely to achieve a significant injection efficiency. Figure 10
depicts a close-up of the density maximum for model v4 at the
same time as in Figure 9, showing that the cloud has been
triggered into roughly spherically symmetric collapse, judging
from the density distribution and the velocity vectors. At the
same time, the color field within the region plotted in Figure 10
is essentially zero, implying that if any shock front material
is to be accreted by the growing protostar, it must occur at
some later phase in its evolution.

 Finally, Figures 11, 12, and 13 show the evolution for model v80L, 
where injection occurred, but dynamical collapse did not ensue.
In this case, the strong shock front tends to shred the target
cloud into streamers (Figure 11), an excellent situation for
injecting shock-front material into the same region (Figure 12),
but not well-suited for inducing sustained dynamic collapse. Model v80L
achieved a maximum density of $\sim 10^{-15}$ g cm$^{-3}$ by the 
time shown in Figure 11 of 0.056 Myr, and by 0.1 Myr (Figure 13), 
the maximum density has dropped to $\sim 10^{-16}$ g cm$^{-3}$. 
Figure 12 also demonstrates that even with the more vigorous
shock compressional heating in model v80L, the molecular cooling
is able to limit the shock-heated regions to the close vicinity
of the shock-cloud boundary.

 In summary, Figure 14 shows how the critical outcomes of 
achieving injection and sustained collapse depend on the 
assumed shock speed, for all the models listed in Table 2. 
Low speed shocks can induce collapse, but not injection, while
high speed shocks result in significant injection, but not in collapse. 
Shocks falling in the range of about 5 km/sec to 70 km/sec appear to
be able to simultaneously induce collapse and achieve injection
of significant amounts of shock wave material. This result is
consistent with the prediction by Foster \& Boss (1996)
that shock speeds of $\sim 100$ km/sec or higher would shred
the target clouds and prevent the formation of a collapsing 
protostar.

 Table 2 also shows that these basic results are relatively
insensitive to the amount of spatial resolution employed,
specifically to the number of AMR grid levels allowed. 
Model v5 with the standard 5 levels of AMR and model v5-4 with 4 levels
resulted in quite similar outcomes, as was the case for
model v20 with 5 levels and model v20-6 with 6 levels.

\subsection{Injection Efficiencies}

 Boss et al. (2008) found that for a 1.0 $M_\odot$ target cloud 
and a 20 km/sec shock, the injection efficiency $f_i$ was 0.003.
For the comparable 2.2 $M_\odot$ target cloud in model v20,
$f_i$ was 0.001, a factor of three times lower for the higher
mass cloud. This difference may be attributed to the fact that
while the shock front was identical in both models, the larger
mass of the target cloud in model v20 made the task of the shock
front more difficult: evidently somewhat lower mass clouds are easier
to trigger into collapse and pollute with shock front material
than somewhat higher mass clouds, at least to the phase studied
by these models (i.e., maximum densities less than 
$\sim 10^{-12}$ g cm$^{-3}$).

 Estimates of the injection efficiency $f_i$ have dropped steadily
as the spatial resolution and physical modeling have improved. Boss (1995) 
found that about half of the impinging shock material ($f_i = 0.5$) 
entered the collapsing cloud in his coarsely-gridded, 3D isothermal
models. Foster \& Boss (1997) found $f_i$ between 0.1 and 0.2 in
their relatively coarsely-gridded, 2D isothermal models, while
Vanhala \& Boss (2000, 2002) found $f_i \sim 0.1$ in their
increasingly higher spatial resolution, 2D isothermal models.

 Broadly speaking, a typical value of $f_i \sim 0.001$ characterizes 
all of the successful triggering and injection models listed in 
Table 2, a value considerably lower than those previously found.
This difference may be attributed to a number of factors, principally
the improved spatial resolution of the current models and the
inclusion of nonisothermal heating effects, both of which appear
to have the effect of reducing $f_i$ compared to lower resolution,
isothermal calculations. The superior shock-handling ability of
the PPM hydrodynamics method that FLASH is based upon undoubtedly
also plays a role. Finally, there is the question of how $f_i$ 
is defined, and when it is evaluated: in Vanhala \& Boss (2002),
e.g., $f_i$ was typically evaluated at earlier times than in
the present models, and the region over which the color field
was considered to have been injected was liberally interpreted
to extend quite some distance from the collapsing protostar
(e.g., Figure 3 in Vanhala \& Boss 2002). When the region
expected to be accreted by the protostar is defined in a
similar manner to that used in the present models, an estimate
of $f_i \sim 0.002$ results from the Vanhala \& Boss (2002) 
models, an estimate more in line with the current values.

 Nevertheless, such a low injection efficiency may still be in 
accord with a supernova as the source of the shock wave. Based on 
the estimates of Cameron et al. (1995), Foster \& Boss (1997) noted 
that the $^{26}$Al-containing gas and dust in a supernova shock 
wave would have to be diluted by a factor of $\sim 10^4$ in order
to explain the inferred initial abundance of $^{26}$Al in the
solar nebula, i.e., $10^{-4} M_\odot$ of supernova shock-wave material
should be injected into a 1 $M_\odot$ presolar cloud. 

 More recently, Takigawa et al. (2008) used detailed models of
a faint supernova with mixing and fallback to attempt to match
the inferred initial abundances of the SLRIs $^{26}$Al, $^{41}$Ca,
$^{53}$Mn, and $^{60}$Fe in the solar nebula, based on  nucleosynthetic 
yield calculations (e.g., Rauscher et al. 2002). They found that a
dilution factor of $D \sim 10^{-4}$ and a time interval of 1 Myr
between the supernova explosion and the formation of the first
refractory solids in the solar nebula was able to do a good job
of matching all four initial abundances. The dilution factor $D$
is the ratio of the amount of mass derived from the supernova
that ends up in the solar nebula to the amount of mass in the solar
nebula that did not derive from the supernova, i.e., the mass derived
from the target cloud in the present models. Takigawa et al. (2008)
found that the best estimates for $D$ depended on the assumed mass
of the pre-supernova star, ranging from $D = 1.3 \times 10^{-4}$
for a 25 $M_\odot$ star to $D = 1.9 \times 10^{-3}$
for a 20 $M_\odot$ star; stars with masses of 30 and 40 $M_\odot$ 
led to intermediate values of $D$. Gaidos et al. (2009), on the
other hand, suggest a value of $D$ of at least $3 \times 10^{-3}$ 
for a 25 $M_\odot$ progenitor star, significantly larger than
the estimate by Takigawa et al. (2008). 

 Trigo-Rodr\'iguez et al. (2009)
have shown that a 6.5 $M_\odot$ AGB star could have produced
the inferred initial abundances of the SLRIs $^{26}$Al, $^{41}$Ca,
$^{60}$Fe, and $^{107}$Pd in the solar nebula, with a similar
dilution factor of $D \sim 3 \times 10^{-3}$. The planetary nebulae
formed by AGB stars typically have slow wind speeds of 10 km/sec,
while the fast winds caused by AGB flashes can overtake the slow
winds and produce swept-up shells with speeds of 30 km/sec
(Frank \& Mellema 1994, their Figures 1 and 2). In fact, the
swept-up shell in the Frank \& Mellema (1994) model closely
resembles the structure assumed for the shock front here and in the standard
cases of Foster \& Boss (1996, 1997): a shock front with a number
density of $\sim 10^4$ cm$^{-3}$ and a thickness of $10^{16}$ cm,
leading to a total shock front mass of 0.016 $M_\odot$ impacting
the target cloud. Based on the present models, then, AGB-derived
shocks moving at speeds of $\sim$ 30 km/sec should be able to
trigger collapse and injection in the same manner as a supernova
shock with similar properties. 

 In the present models, the mass of the shock wave 
that is incident on the target cloud is 0.016 $M_\odot$, so
values of $f_i \sim 0.001$ imply that about $2 \times 10^{-5} M_\odot$
of shock front material is injected into the collapsing
protostar. If the final result is a well-mixed $\sim 1 M_\odot$ 
protostar and protoplanetary disk, then the dilution factor $D$
is $\sim 2 \times 10^{-5}$, considerably lower than required
by either Takigawa et al. (2008) or Gaidos et al. (2009).
In fact, the mismatch is even worse than this, because in
the case of a supernova shock front, the material ejected by
the supernova must be diluted by the swept-up, intervening
interstellar medium that is necessary to slow down the shock
front to speeds capable of achieving triggering and injection.
A supernova shock launched at 1000 km/sec must snowplow at
least 15 times more mass in order to slow below 70 km/sec.
Hence the dilution factor for a supernova shock must be
decreased by this same factor, to $D \sim 10^{-6}$.

 Clearly there is a need to learn if these crudely estimated
dilution factors can be increased to values closer to those 
advocated by Takigawa et al. (2008) or Gaidos et al. (2009).
Not all of the 2.2 $M_\odot$ target cloud will be
accreted by the growing protostar, so the injection efficiency 
will be larger by a proportionate amount. In addition, if the
bulk of the shock front material infalls somewhat later than
the earliest arrivals (as suggested by the color waves in
Figure 7), then the shock front material may end up preferentially
in the protoplanetary disk, rather than in the star, thereby
increasing proportionally the dilution factor in the disk. Since 
the present models do not include rotation of the target cloud
or shock front, there is no possibility for a rotationally
supported disk to form, and so the present models cannot fully
answer the question of the dilution factor appropriate for the 
solar nebula, as opposed to the presolar cloud as a whole. Calculations
are currently underway on the dc101 cluster at DTM that include 
rotation for 2D target clouds, in order to address this key question.

 Given the lower injection efficiency for a 2.2 $M_\odot$
cloud compared to a 1.0 $M_\odot$ cloud, higher cloud densities
evidently result in lower injection efficiencies, as might
be expected. Target clouds with lower initial densities should
then have higher injection efficiencies, and their larger radii
(for a given mass cloud) will also go in the direction of
increasing the total amount of injected shock wave material.
Furthermore, allowing injection of shock wave material from behind
the leading edge of the shock front (here considered to be
only 0.003 pc thick) will also increase the amount injected.
Increasing the assumed density and thickness of the shock front should also
lead to higher injection efficiencies. Future work will include the study
of shock fronts with different densities and thicknesses compared
to the standard case of Foster \& Boss (1996, 1997) and employed
in the present models, as well as different density and radii target 
clouds, in order to better ascertain the suitability
of a wider range of possible shock fronts and target clouds
for triggering and injection.
 
 Coupled with these concerns over the low injection efficiency and the
dilution factor are the implications for the star-formation environment
where simultaneous triggering and injection might have occurred.
Looney, Tobin, \& Fields (2006) showed that for the target cloud
considered here and an injection efficiency of $f_i = 0.1$, due to
geometric dilution alone, a supernova would have to occur within about 
0.06 pc to 1.2 pc from the target cloud core in order to inject the 
desired amount of SLRIs. For the much lower values of $f_i$ found here,
$f_i \sim 0.001$, these distance estimates decrease by factors of 10,
to 0.006 pc to 0.12 pc. Such distances are appropriate for the proplyds 
in the Orion nebula that are being photoevaporated by the Orion Trapezium's 
four O stars (e.g., Williams, Andrews, \& Wilner 2005). However,
the proplyds have already collapsed to form protostars, and the density 
of the surrounding HII region is too low to slow down a supernova shock 
wave by snowplowing to the required speeds. Thus it remains to be seen 
if a combination of target cloud and shock front parameters can
be found that will increase the injection efficiencies sufficiently
to produce a scenario that is consistent with observations of regions
of high mass star formation (e.g., Hester \& Desch 2005).

\section{Conclusions}

 When cooling by appropriate molecular species is included, shocks with 
speeds in the range from 5 km/sec to 70 km/sec are able to trigger the
gravitational collapse of otherwise stable, dense cloud cores,
as well as to inject shock wave
material into the collapsing cloud cores. This injected material consists
of shock wave gas as well as dust grains small enough to remain coupled to 
the gas, i.e., sub-micron-sized grains, which are expected to characterize 
supernova shock waves (Bianchi \& Schneider 2007) and to carry the
SLRI whose decay products have been found in refractory inclusions of
chondritic meteorites. Evidently a radiative-phase supernova shock wave 
(Chevalier 1974) is able to cool sufficiently rapidly to behave in much
the same way as a shock wave that is assumed to remain isothermal with 
the target cloud (e.g., Boss 1995). Given that Wolf-Rayet star winds 
and supernova shocks both are launched with shock speeds on the order 
of $10^3$ km/sec, these shock waves can only trigger collapse after 
they have travelled some distance (typically about 10 pc) and been slowed
down to 5 km/sec to 70 km/sec by the snowplowing of intervening interstellar
cloud gas and dust. The distance a fast shock must travel in order to slow 
down to speeds consistent with simultaneous triggering and injection is
inversely proportional to the mean density of the intervening material
(assuming this material to be moving much less than the shock speed);
typical interstellar medium densities lead to distances of a few pc,
depending on the desired shock speed. An AGB star wind with a typical 
speed of 10 to 30 km/sec could also have triggered collapse and injection
without the need for snowplowing. The low injection efficiencies of
the present models, however, point to the need to consider shock
fronts and target clouds with different parameters, in order to
learn if the injection efficiencies can be increased to the levels
thought to be necessary to explain the observed abundances of fossil
SLRIs in meteorites.

 We are currently running three-dimensional (3D) models on the Xenia 
cluster at DTM. The need for a 3D treatment of the shock triggering 
process is evident from previous 3D studies of the R-T instability
(Stone \& Gardiner 2007) and shock fronts (Stone \& Norman 1992;
Whalen \& Norman 2008). The R-T ``sheets'' that form in axisymmetric 
2D models become true R-T fingers in 3D, allowing better penetration 
into the target dense cloud cores. We will present the results of 
these ongoing 3D models as well as of models with varied shock fronts
and target clouds in future papers.

\acknowledgements

 We thank the referee, Steven Desch, for several constructive reports 
that have led to a number of significant improvements in this paper.
The calculations were performed primarily on the dc101 cluster at DTM.
This research was supported in part by NASA Origins of Solar Systems 
grant NNG05GI10G and NASA Planetary Geology and Geophysics grant 
NNX07AP46G, and is contributed in part to NASA Astrobiology Institute
grant NCC2-1056. The software used in this work was in part
developed by the DOE-supported ASC/Alliances Center for 
Astrophysical Thermonuclear Flashes at the University of Chicago.

\clearpage
\begin{deluxetable}{lcccccc}
\tablecaption{Comparisons with the standard case of Foster \& Boss (1996).
\label{tbl-1}}
\tablewidth{0pt}
\tablehead{\colhead{Model} 
& \colhead{$N_{BR}$} 
& \colhead{$N_{BZ}$} 
& \colhead{$N_{L}$} 
& \colhead{$\rho_{max}$} 
& \colhead{$f_i$} 
& \colhead{$t_f$} }
\startdata

FBA & 5 &  5 & 5 & 5.3e-13 & 0.005 & 5.0e12 \\

FBB & 5 & 10 & 5 & 1.7e-12 & 0.003 & 4.8e12 \\

FBC & 5 & 10 & 6 & 8.0e-12 & 0.002 & 4.7e12 \\

FBD & 5 & 15 & 5 & 2.0e-12 & 0.002 & 4.3e12 \\

\enddata
\end{deluxetable}

\clearpage

\begin{deluxetable}{lccccccc}
\tablecaption{Nonisothermal models with varied shock speeds.
\label{tbl-2}}
\tablewidth{0pt}
\tablehead{\colhead{Model} 
& \colhead{$v_s$} 
& \colhead{$N_{BR}$} 
& \colhead{$N_{BZ}$} 
& \colhead{$N_{L}$} 
& \colhead{$\rho_{max}$} 
& \colhead{$f_i$} 
& \colhead{$t_f$} }
\startdata

v1    & 1.0  & 5 & 15 & 5 & 5.e-12 & 0.0    & 3.8e12 \\

v2    & 2.0  & 5 & 15 & 5 & 1.e-12 & 0.0    & 8.5e12 \\

v2.5  & 2.5  & 5 & 15 & 5 & 2.e-12 & 0.0    & 4.8e12 \\

v4-4  & 4.0  & 5 & 15 & 4 & 1.e-12 & 0.0    & 1.2e12 \\

v4    & 4.0  & 5 & 15 & 5 & 5.e-12 & 0.0    & 3.6e12 \\

v5-4  & 5.0  & 5 & 15 & 4 & 2.e-12 & 2.e-4  & 7.5e12 \\

v5    & 5.0  & 5 & 15 & 5 & 1.e-12 & 3.e-4  & 8.5e12 \\

v7    & 7.0  & 5 & 15 & 5 & 2.e-12 & 6.e-4 & 8.5e12 \\

v9    & 9.0  & 5 & 15 & 5 & 1.e-12 & 2.e-4 & 1.2e12 \\

v10   & 10.0 & 5 & 15 & 5 & 5.e-12 & 2.e-3 & 6.5e12 \\

v20   & 20.0 & 5 & 15 & 5 & 1.e-12 & 1.e-3 & 5.3e12 \\

v20-6 & 20.0 & 5 & 15 & 6 & 1.e-12 & 4.e-4 & 5.2e12 \\

v30   & 30.0 & 5 & 15 & 5 & 4.e-12 & 3.e-3 & 4.0e12 \\

v40   & 40.0 & 5 & 15 & 5 & 2.e-12 & 1.e-3 & 1.2e13 \\

v50   & 50.0 & 5 & 15 & 5 & 1.e-12 & 4.e-4 & 4.7e12 \\

v60   & 60.0 & 5 & 15 & 5 & 1.e-12 & 4.e-4 & 6.5e12 \\

v70   & 70.0 & 5 & 15 & 5 & 1.e-12 & 3.e-4 & 6.5e12 \\

v75   & 75.0 & 5 & 20 & 5 & 3.e-13 & 3.e-4 & 8.5e12 \\

v75L  & 75.0 & 5 & 20 & 5 & 1.e-15 & 6.e-4 & 4.5e12 \\

v80L  & 80.0 & 5 & 20 & 5 & 1.e-15 & 5.e-4 & 3.7e12 \\

v90L  & 90.0 & 5 & 20 & 5 & 1.e-15 & 4.e-4 & 3.4e12 \\

v100L &100.0 & 5 & 20 & 5 & 1.e-15 & 6.e-4 & 3.0e12 \\

\enddata
\end{deluxetable}
\clearpage

\begin{figure}
\vspace{-1.0in}
\plotone{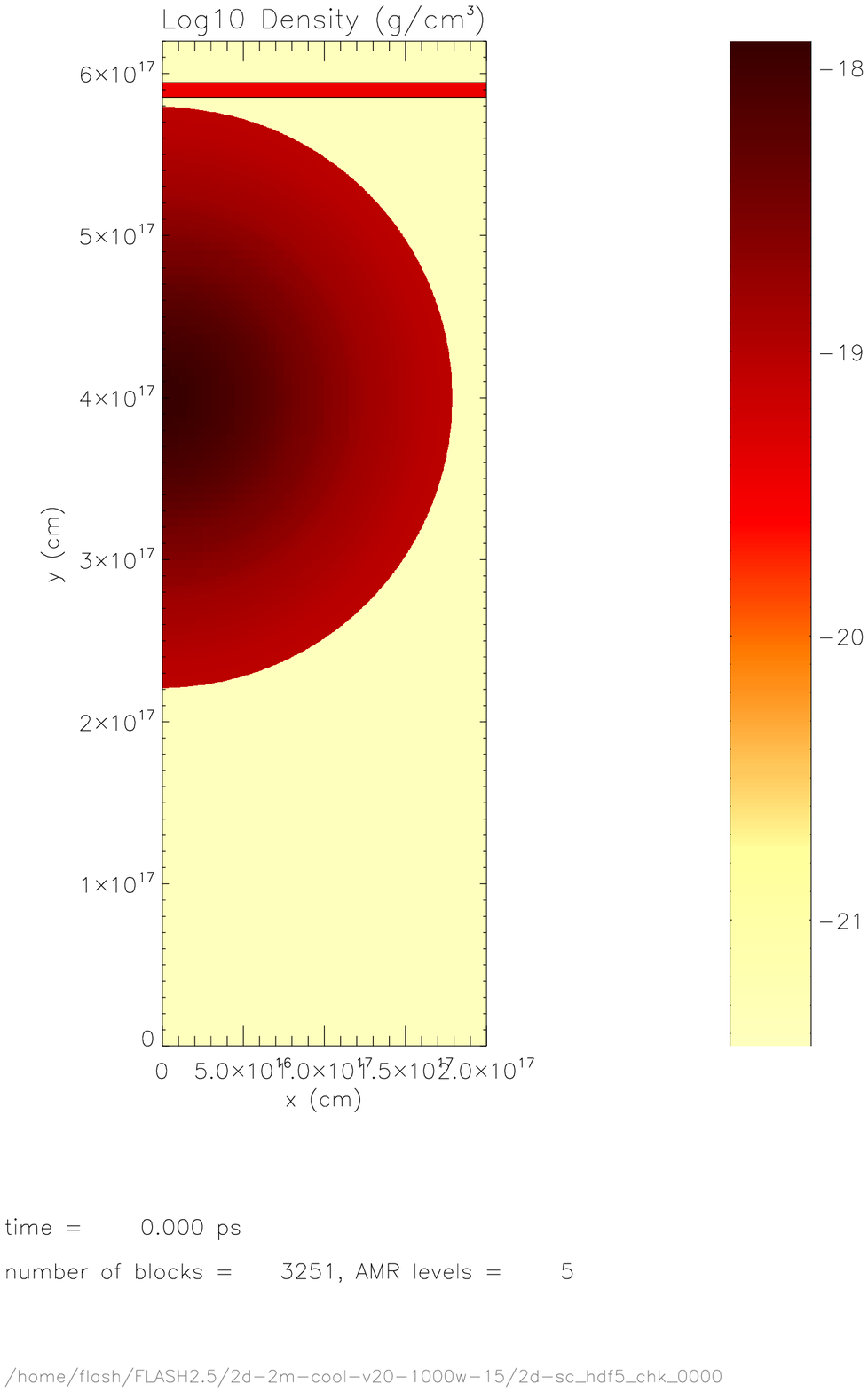}
\vspace{+0.25in}
\caption{Initial log density distibution for all the nonisothermal
models (Table 2) with varied shock speeds. Black contours show 
regions with color fields (representing SLRI) greater than 0.001
(dimensionless units) within the shock wave, which is moving downward 
from the top of the box and is about to strike the target cloud.
The symmetry axis is along the left hand side of the plot. 
The $R$ axis is horizontal and the $Z$ axis is vertical.}
\end{figure}

\begin{figure}
\vspace{-1.0in}
\plotone{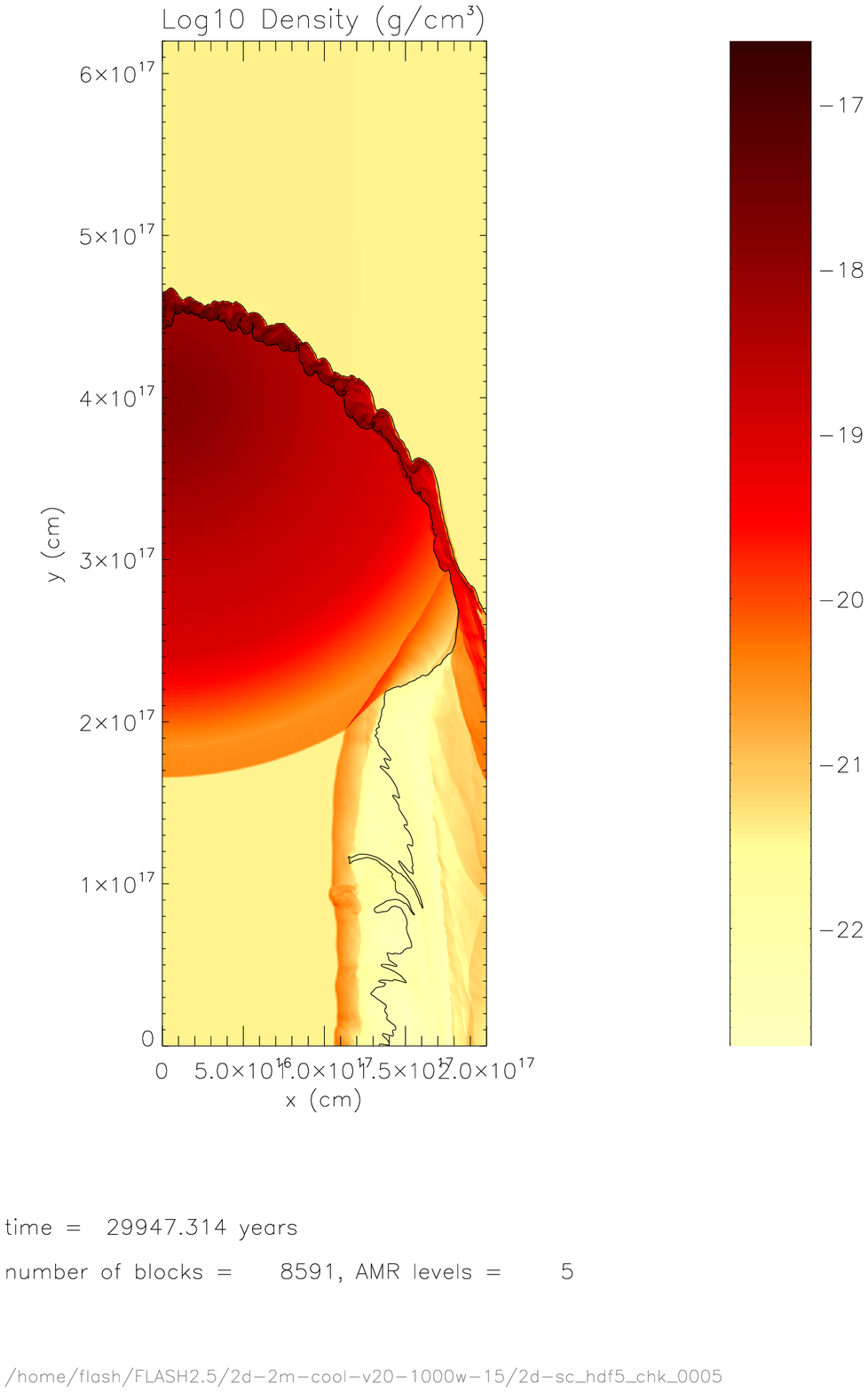}
\vspace{+0.25in}
\caption{Model v20 after 29,947 yr, plotted in the same manner as
in Figure 1. R-T fingers and K-H vortices have formed at the shock-cloud 
interface, simultaneously injecting some shock wave material into the
target cloud while ablating other portions of the cloud into 
the downstream flow.}
\end{figure}

\begin{figure}
\vspace{-1.0in}
\plotone{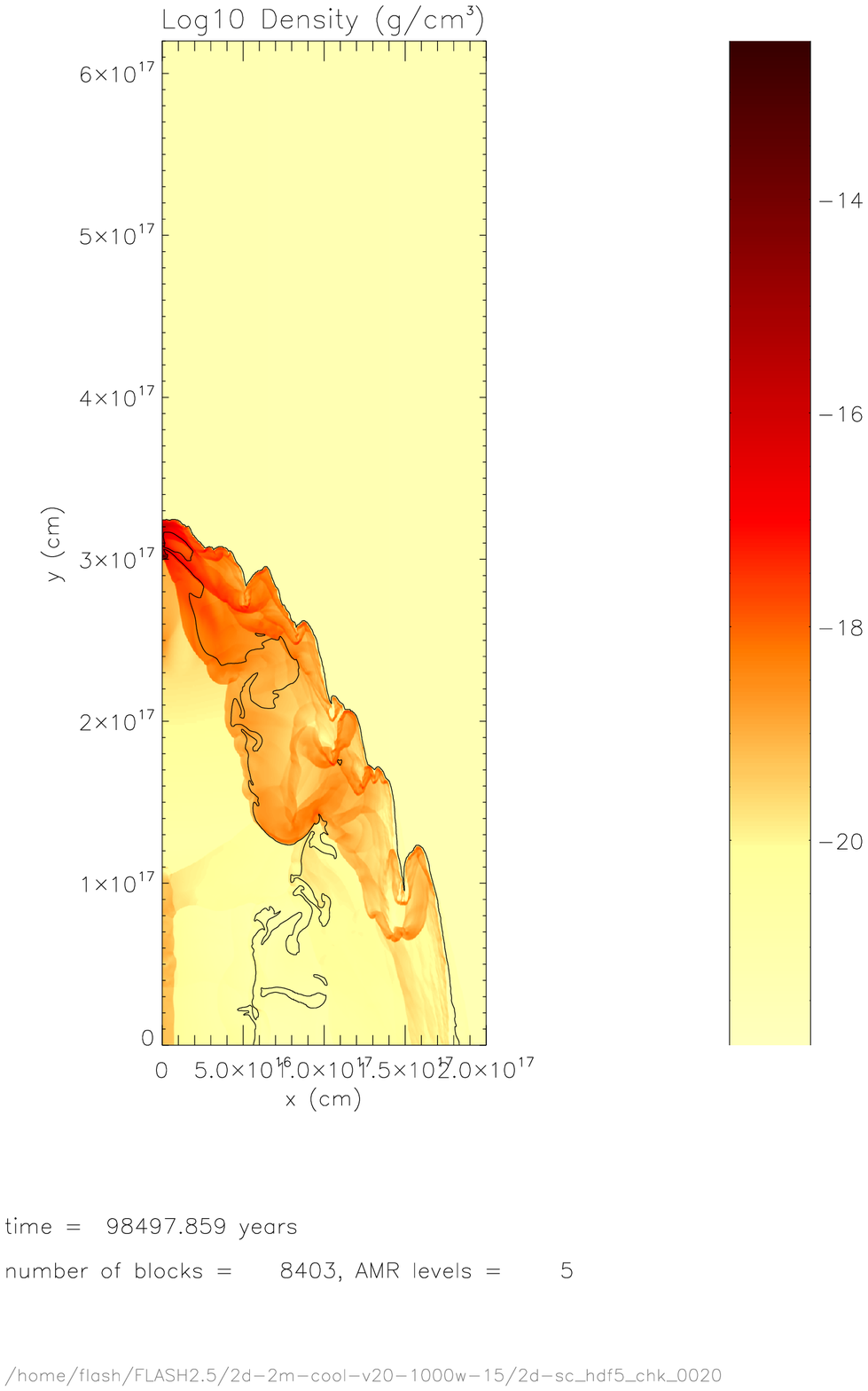}
\vspace{+0.25in}
\caption{Model v20 after 98,498 yr, showing the formation of a dense,
dynamically collapsing protostar on the symmetry axis at the middle
of the box.}
\end{figure}

\begin{figure}
\vspace{-1.0in}
\plotone{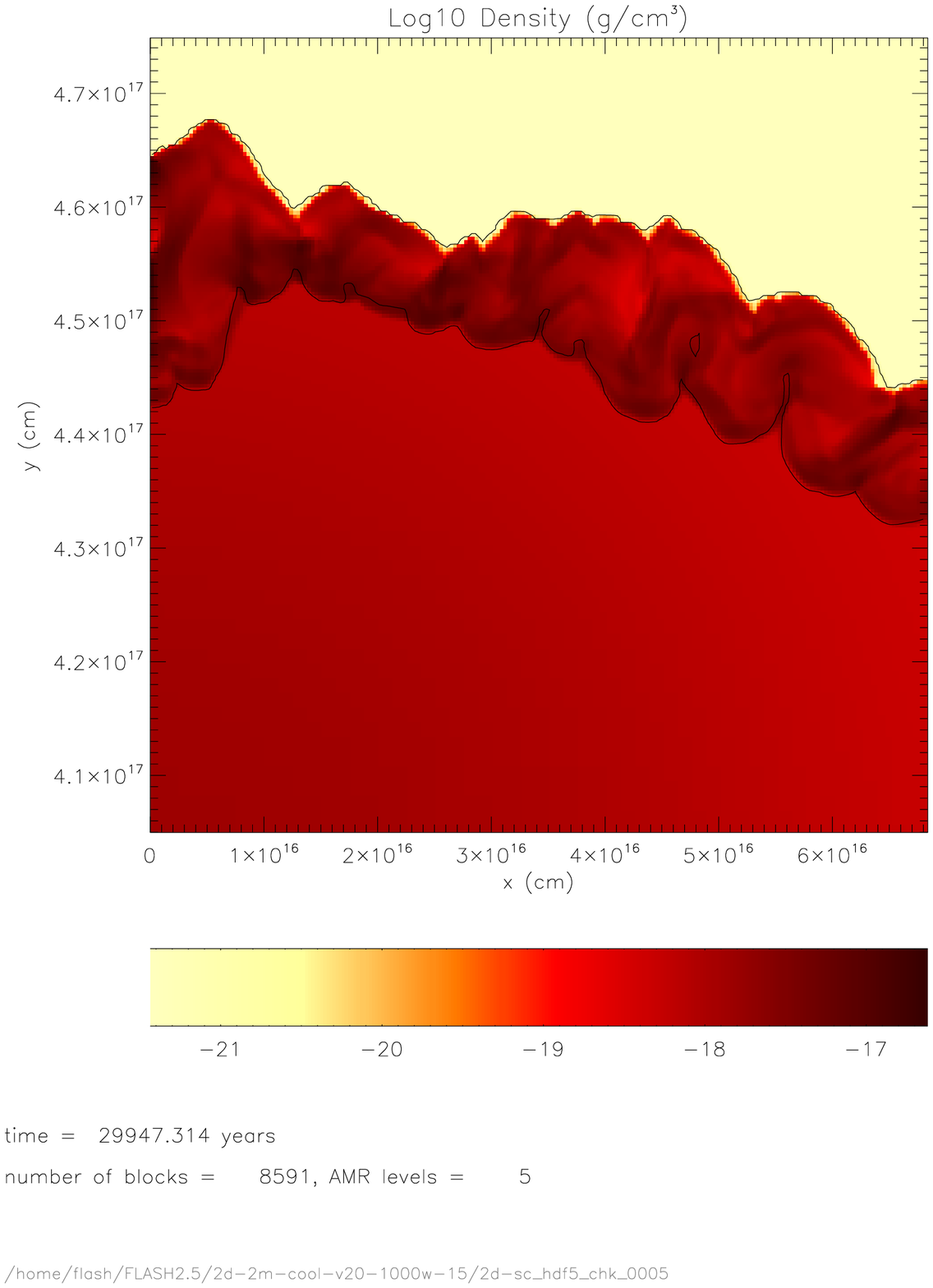}
\vspace{+0.25in}
\caption{Model v20 after 29,947 yr, showing the region around the shock 
front. The R-T fingers and K-H vortices contain the shock front material, 
as they lie within the black contour lines for the color field.}
\end{figure}

\begin{figure}
\vspace{-1.0in}
\plotone{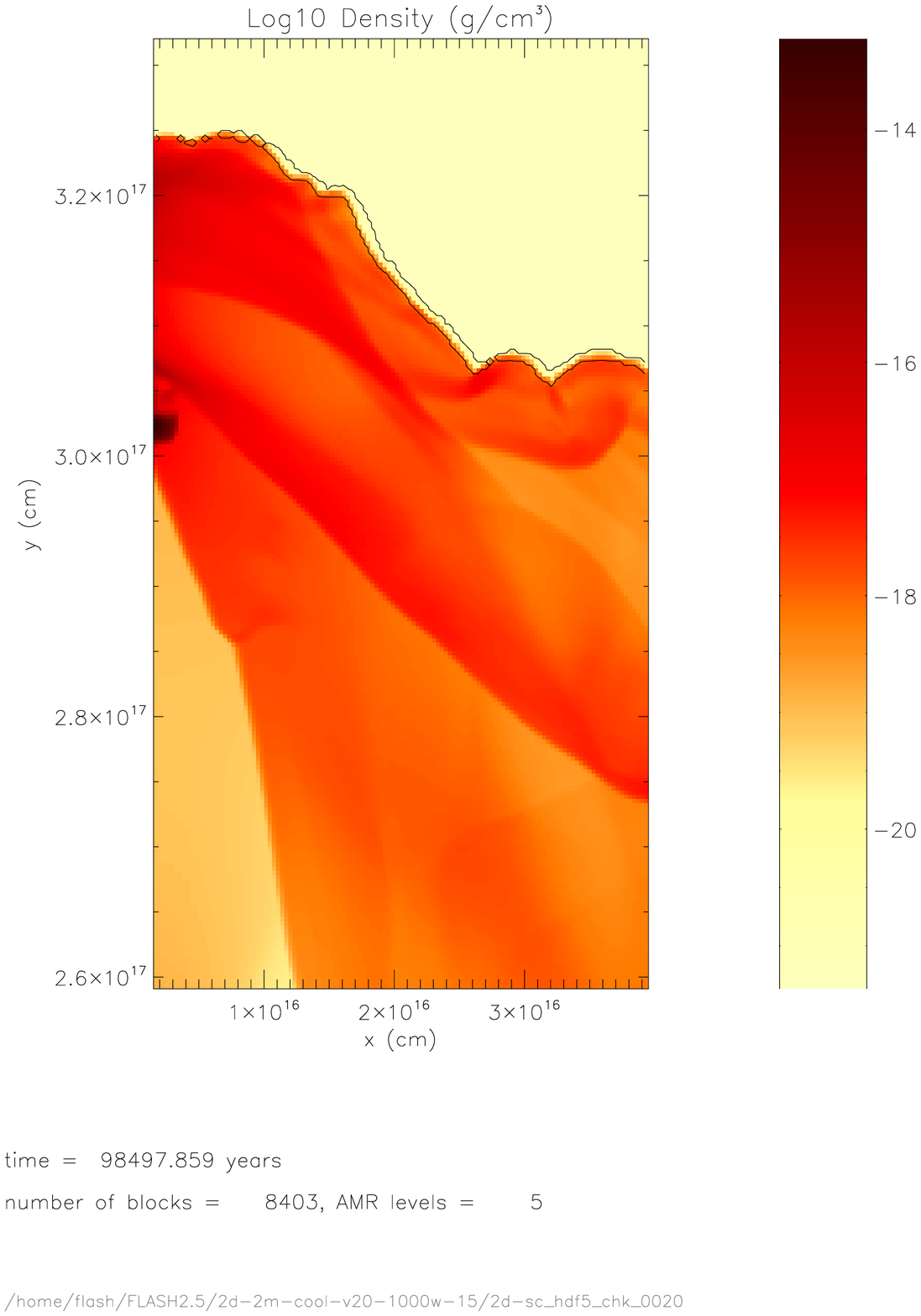}
\vspace{+0.25in}
\caption{Model v20 after 98,498 yr, showing the region around
the collapsing protostar. The black contours now
show regions with temperatures greater than 100 K, which only
occur at the shock-cloud interface as a result of the molecular
cooling. A high-density region, the protostar, has formed 
along the symmetry axis.}
\end{figure}

\begin{figure}
\vspace{-1.0in}
\plotone{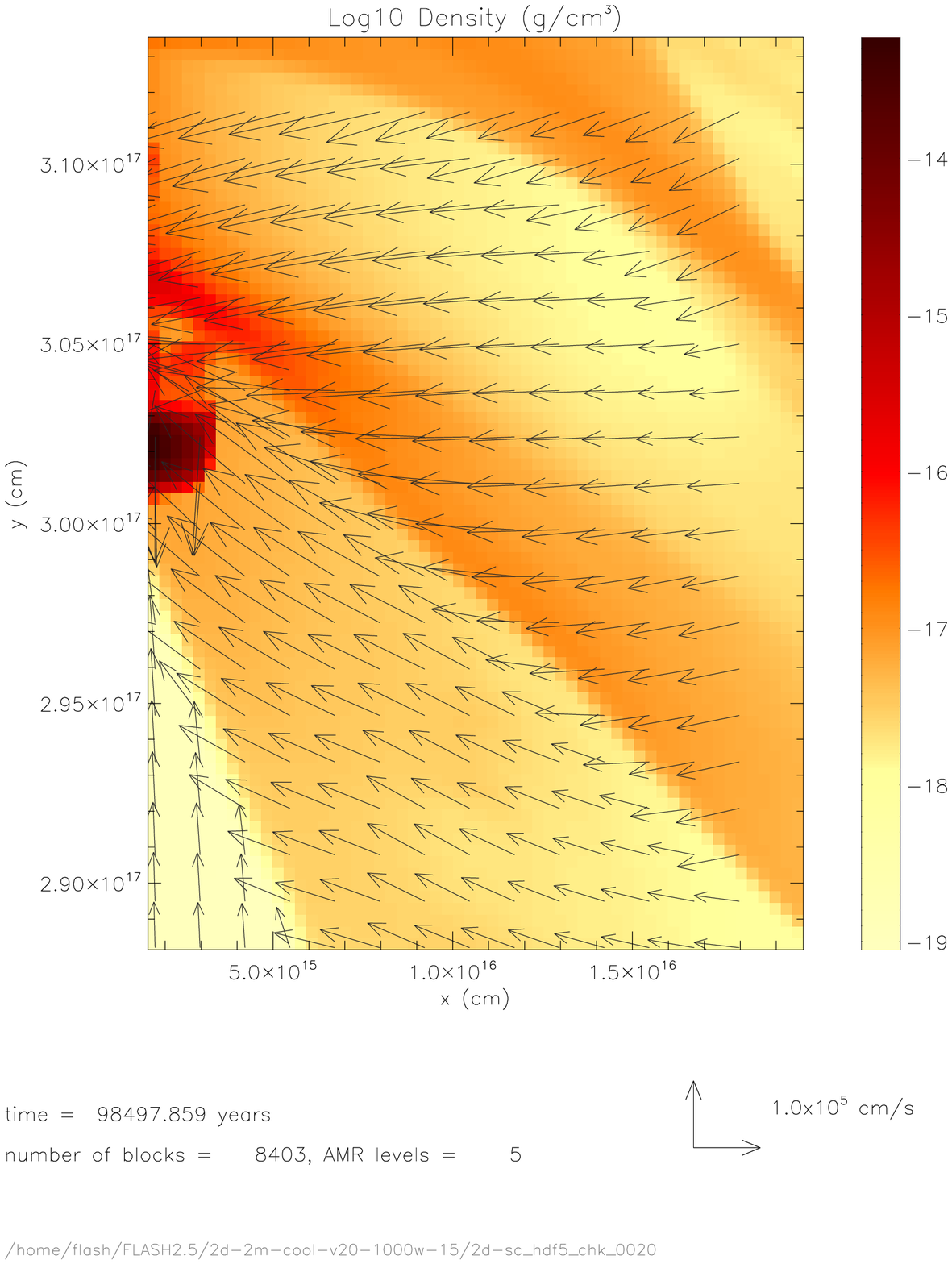}
\vspace{+0.25in}
\caption{Model v20 after 98,498 yr, as in Figure 5, except limited to a 
small region around the density maximum of $\sim 10^{-13}$ g cm$^{-3}$.
Velocity vectors are shown for every fourth AMR grid cell in 
$R$ and $Z$; their scale bar is 1 km/sec. The collapsing
protostar has been accelerated by the shock front into downward
motion at a speed of $\sim$ 1 km/sec, while much of the rest of the
cloud envelope is infalling toward the growing protostar.}
\end{figure}

\begin{figure}
\vspace{-1.0in}
\plotone{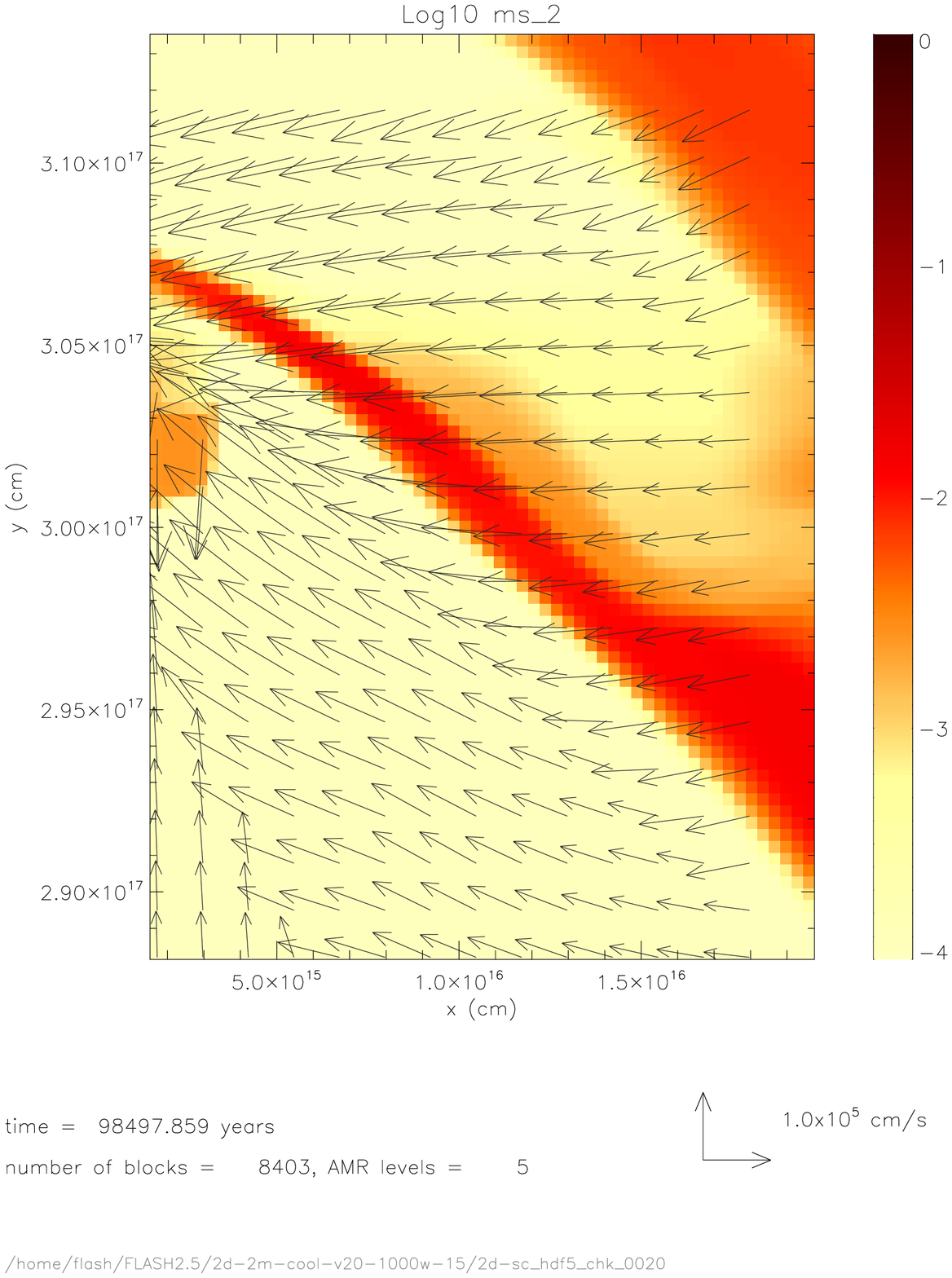}
\vspace{+0.25in}
\caption{Same as Figure 6 for model v20, except now the log of 
the color field is plotted, showing that the collapsing protostar
has been injected with significant material derived from the
shock front, i.e. SLRI. Several waves of color-rich material
appear to be headed toward the protostar.}
\end{figure}

\begin{figure}
\vspace{-1.0in}
\plotone{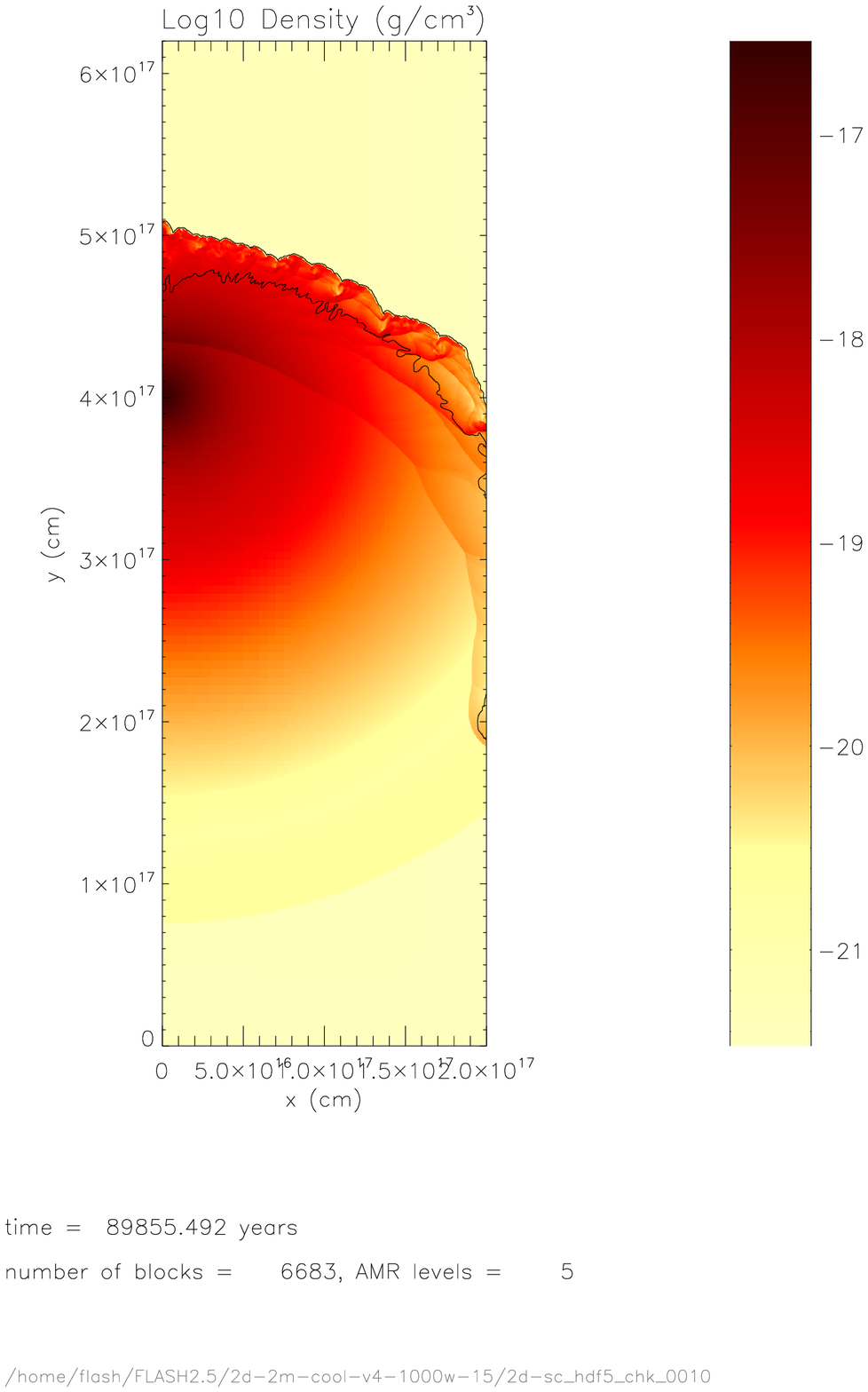}
\vspace{+0.25in}
\caption{Density distribution for model v4 after 89,855 yr of evolution, 
plotted as in Figure 1, with regions where the color field is greater 
than 0.001 being denoted by the black contour lines.}
\end{figure}

\begin{figure}
\vspace{-1.0in}
\plotone{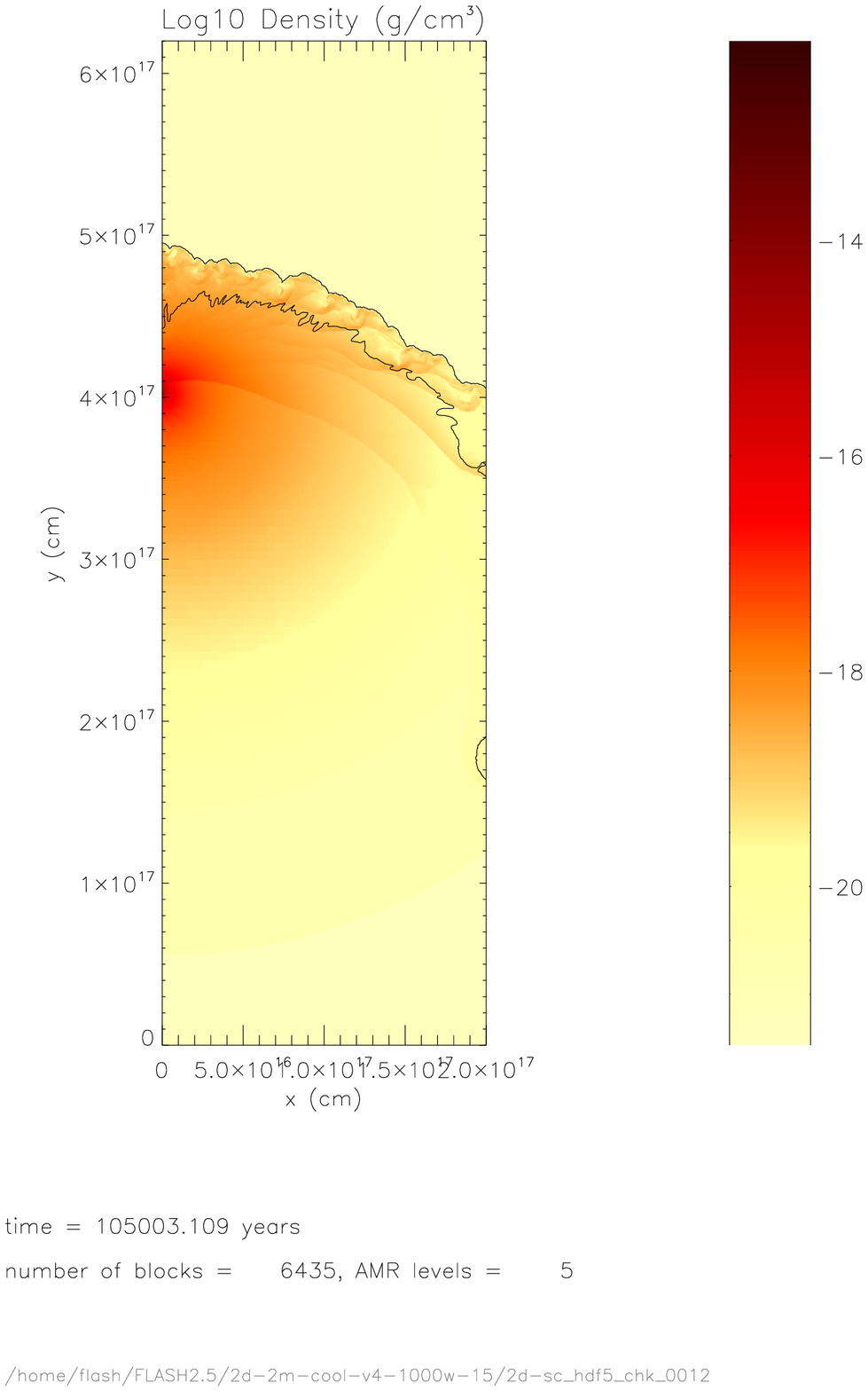}
\vspace{+0.25in}
\caption{Model v4 after 105,003 yr. Compared to Figure 3
for model v20, it is clear that the shock front material has not been
able to penetrate into the densest regions of the collapsing target
cloud, though some of the color field might be injected at later
times.}
\end{figure}

\begin{figure}
\vspace{-1.0in}
\plotone{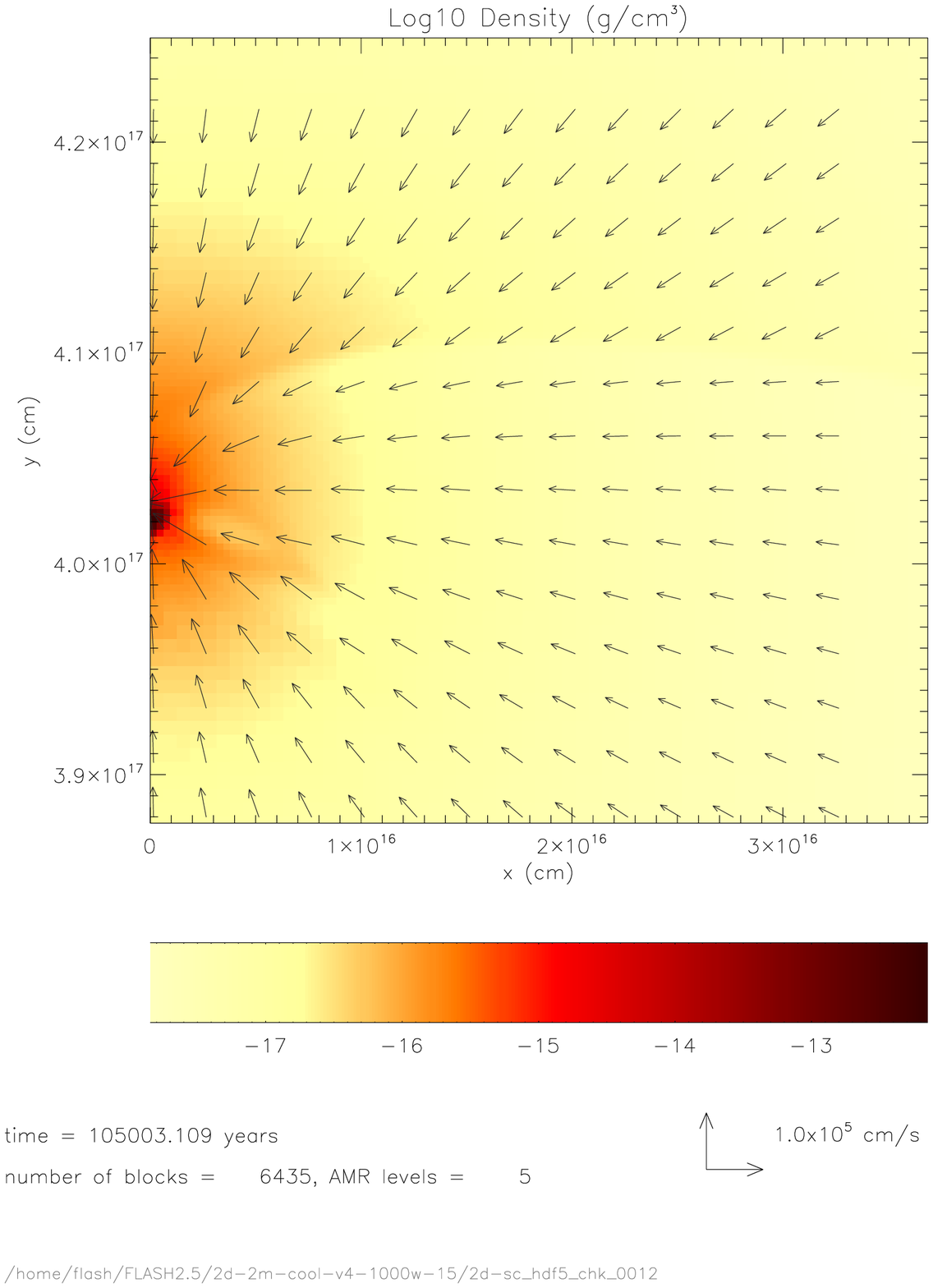}
\vspace{+0.25in}
\caption{Model v4 after 105,003 yr, showing the region around
the collapsing protostar, with a density maximum of 
$\sim 10^{-12}$ g cm$^{-3}$. The color field is effectively zero
throughout this region. Velocity vectors are plotted for every 
eighth AMR grid cell in $R$ and $Z$. The protostar is
collapsing but has not been accelerated downward to a speed of
$\sim 1$ km/sec, as happened with model v20 (Figure 6).}
\end{figure}

\begin{figure}
\vspace{-1.0in}
\plotone{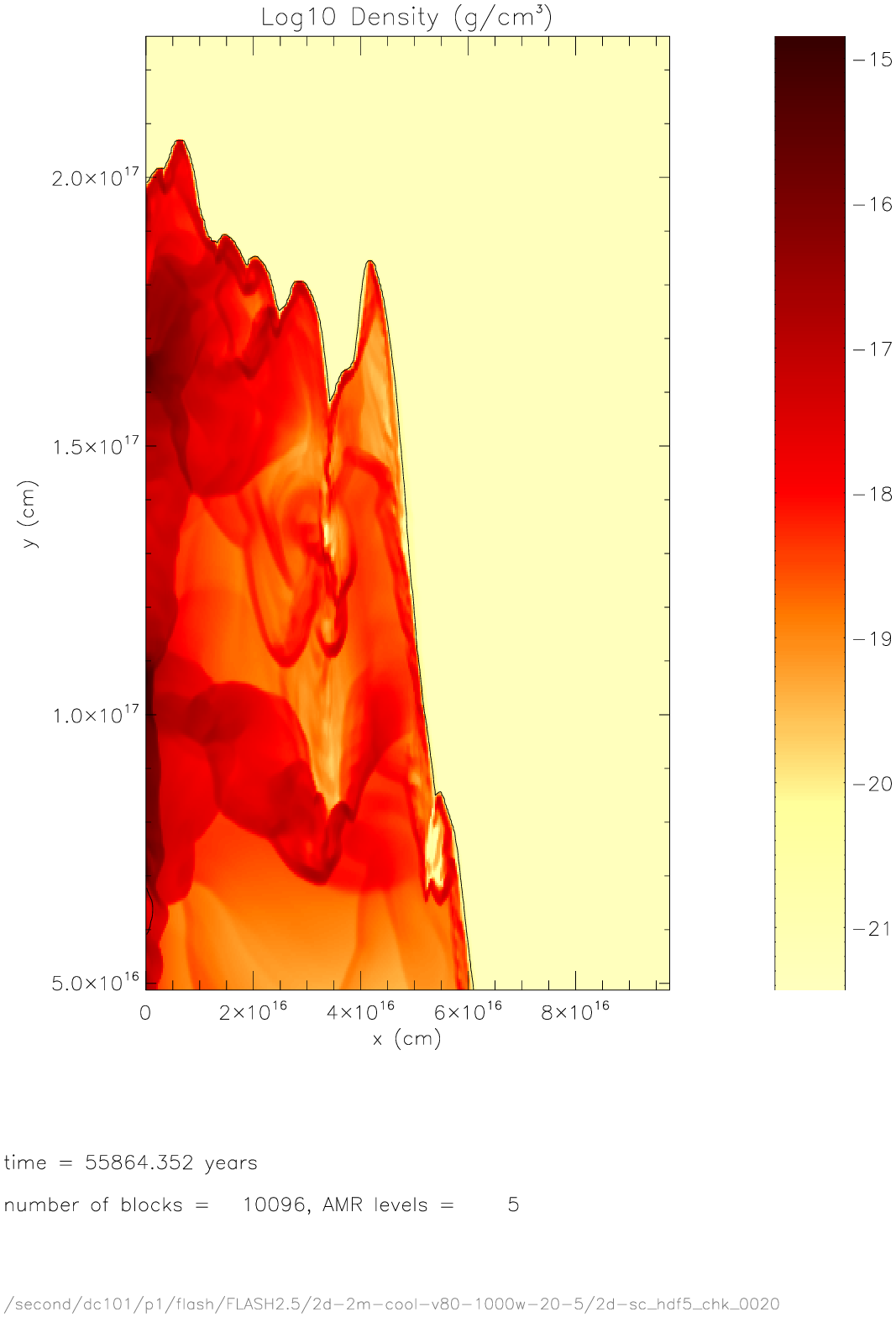}
\vspace{+0.25in}
\caption{Density distribution for model v80 after 55,864 yr of evolution, 
plotted as in Figure 1, but only for the highest
density regions. The target cloud has a much more turbulent structure 
after being struck with this higher speed shock, compared to the
previous models v20 and v4. The density maximum is only 
$\sim 10^{-15}$ g cm$^{-3}$.}
\end{figure}

\begin{figure}
\vspace{-1.0in}
\plotone{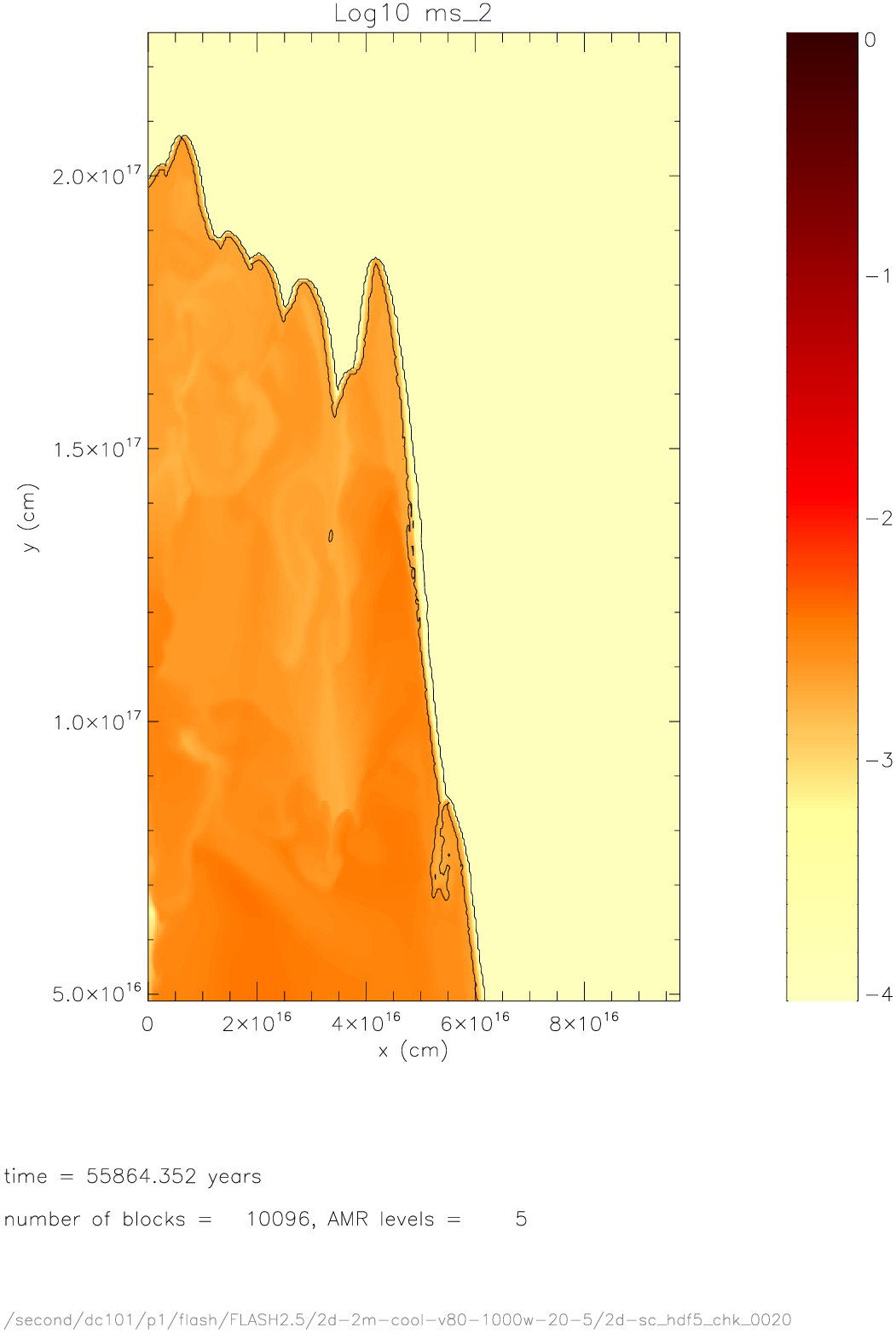}
\vspace{+0.25in}
\caption{Model v80 after 55,864 yr, showing the same region as in
Figure 11, but plotting the log of the color field and temperature 
contours (black) for regions with $T > 100$K. The entire region 
is polluted with shock wave material. Nonisothermal temperatures 
are again limited to the edges of the shock-cloud interface.}
\end{figure}

\begin{figure}
\vspace{-1.0in}
\plotone{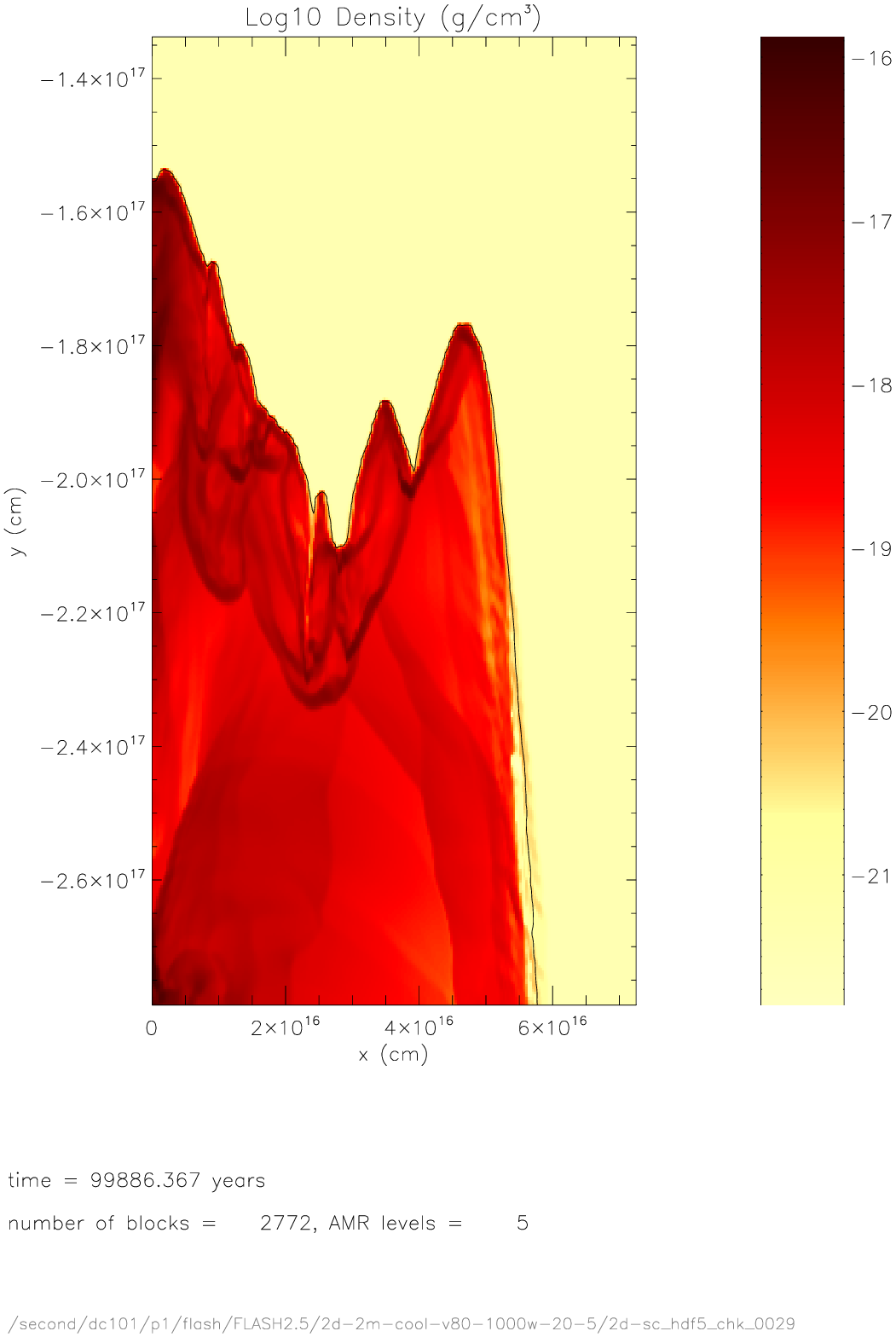}
\vspace{+0.25in}
\caption{Same as Figure 11 after 99,886 yr for model v80. 
The density maximum has dropped to $\sim 10^{-16}$ g cm$^{-3}$;
dynamic collapse leading to protostellar formation has not occurred.}
\end{figure}

\begin{figure}
\vspace{-1.0in}
\plotone{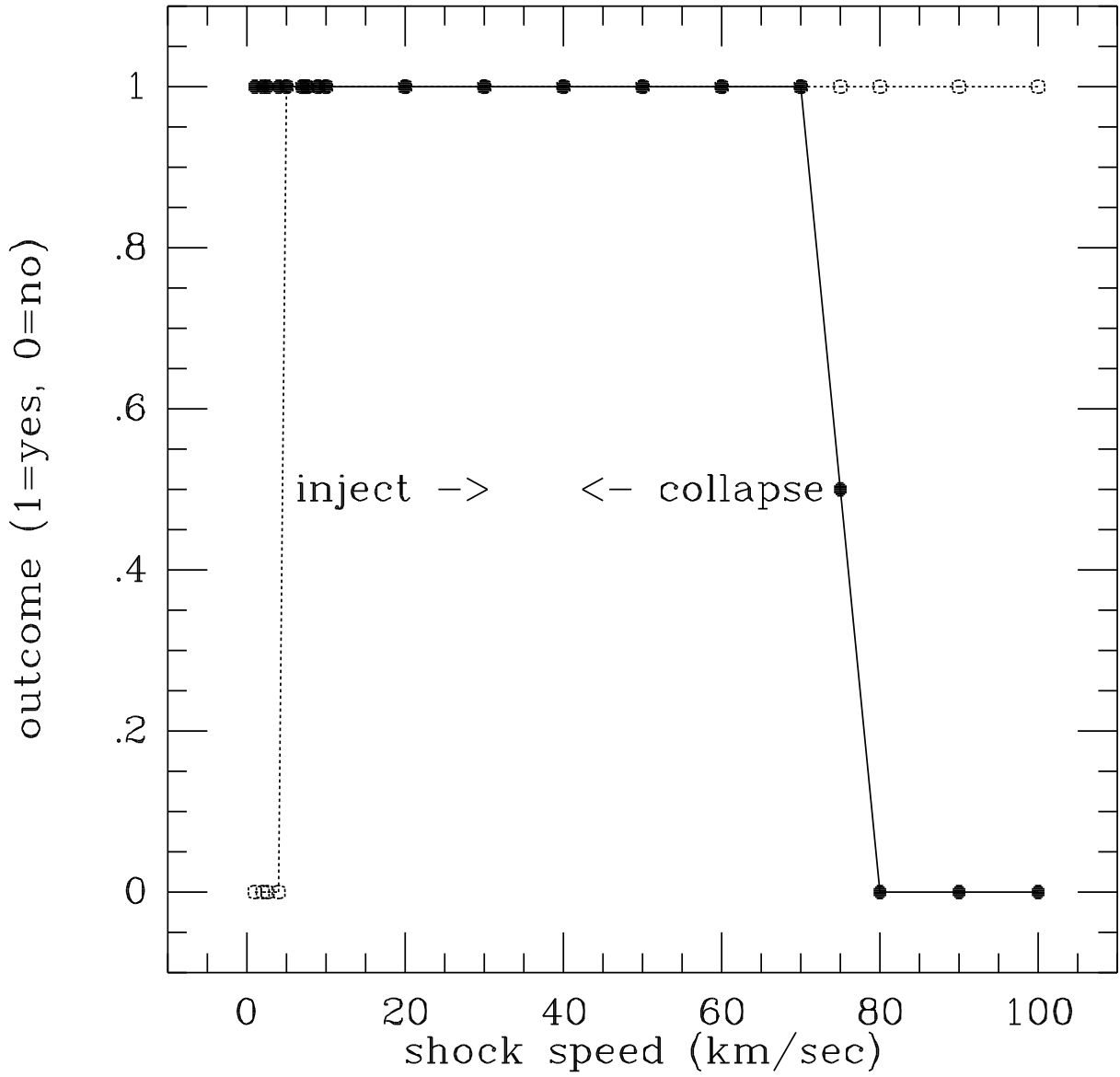}
\vspace{-0.5in}
\caption{Results for the models with varied shock speeds, indicating
whether dynamic collapse resulted (filled circles) or whether shock wave 
material was injected significantly into the dense cloud core (open circles).
The overlap of these two criteria represent successful models for
shock-triggered collapse and injection.}
\end{figure}

\end{document}